\documentclass[aps,10pt,twocolumn,superscriptaddress, nofootinbib]{revtex4-2}
\usepackage{mathrsfs, amssymb, amsmath}  
\usepackage{epsfig, cancel}
\usepackage{footmisc}
\usepackage{latexsym}
\usepackage{natbib}
\usepackage{url}
\usepackage{dcolumn}
\usepackage{multirow}
\usepackage{color}
\usepackage{cancel}
\usepackage{soul}
\usepackage[normalem]{ulem}
\usepackage{amsfonts,amssymb,amsmath, txfonts}
\usepackage{graphicx,epsfig}
\usepackage{psfrag}
\usepackage{hyperref}
\usepackage{mathtools}
\usepackage{enumitem}
\usepackage{float}
\usepackage[dvipsnames]{xcolor}
\usepackage{xcolor}
\hypersetup{linktoc=all,
    colorlinks=true, linkcolor={brightpink},
    citecolor={blue}, urlcolor={blue}
}
\definecolor{rosy}{RGB}{230,235,252}
\definecolor{myframetitle}{RGB}{90,89,170}
\definecolor{myblocktitle}{RGB}{140,185,249}
\definecolor{mytitle}{RGB}{10,80,26}

\definecolor{darkgreen}{RGB}{27,130,45}
\definecolor{darkblue}{rgb}{0,0,0.3}
\definecolor{darkred}{rgb}{0.7,0,0}

\definecolor{light gray}{RGB}{220,220,220}
\definecolor{dark purple}{RGB}{108,0,217}
\definecolor{pink}{RGB}{190,20,100}
\definecolor{orang}{RGB}{193,63,0}
\definecolor{green}{RGB}{11,98,17}
\definecolor{darkpink}{RGB}{153,0,76}
\definecolor{bluegreen}{RGB}{0,102,102}
\definecolor{greenlagan}{RGB}{0,102,0}
\definecolor{redgreen}{RGB}{102,102,0}
\definecolor{Redgreen}{RGB}{153,76,0}
\definecolor{vividviolet}{rgb}{0.62, 0.0, 1.0}
\definecolor{amaranth}{rgb}{0.9, 0.17, 0.31}
\definecolor{palatinateblue}{rgb}{0.15, 0.23, 0.89}
\definecolor{brightpink}{rgb}{1.0, 0.0, 0.5}
\definecolor{cornflowerblue}{rgb}{0.39, 0.58, 0.93}
\definecolor{deepcarminepink}{rgb}{0.94, 0.19, 0.22}
\definecolor{radicalred}{rgb}{1.0, 0.21, 0.37}

\def\H0{{\text{H}\hspace*{-2.05mm}\text{H} 0\hspace*{-1.35mm}0\ }}

\def\be{\begin{equation}}
\def\ee{\end{equation}}
\def\beq{\begin{equation}}
\def\eeq{\end{equation}}
\def\bea{\begin{eqnarray}}
\def\eea{\end{eqnarray}}
\newcommand{\dd}{\textrm{d}}

\begin{document}

\title{Putting Flat $\Lambda$CDM In The (Redshift) Bin}

\author{E. \'O Colg\'ain}
\affiliation{CQUeST \& Department of Physics, Sogang University, Seoul 121-742, Korea}
\affiliation{Atlantic Technological University, Ash Lane, Sligo, Ireland}
\author{M. M. Sheikh-Jabbari}
\affiliation{School of Physics, Institute for Research in Fundamental Sciences (IPM),\\ P.O.Box 19395-5531, Tehran, Iran}
\affiliation{The Abdus Salam ICTP, Strada Costiera 11, I-34014, Trieste, Italy}
\author{R. Solomon}
\affiliation{HEPCOS, Department  of  Physics,  SUNY  at  Buffalo,  Buffalo,  NY  14260-1500, USA}
\author{M. G. Dainotti}
\affiliation{National Astronomical Observatory of Japan, 2 Chome-21-1 Osawa, Mitaka, Tokyo 181-8588, Japan}
\affiliation{The Graduate University for Advanced Studies, SOKENDAI, Shonankokusaimura, Hayama, Miura District, Kanagawa 240-0193, Japan}
\affiliation{Space Science Institute, Boulder, CO, USA}
\author{D. Stojkovic}
\affiliation{HEPCOS, Department  of  Physics,  SUNY  at  Buffalo,  Buffalo,  NY  14260-1500, USA}

\begin{abstract}
Flat $\Lambda$CDM cosmology is specified by two constant fitting parameters at the background level in the late Universe, the Hubble constant $H_0$ and matter density (today) $\Omega_m$. Mathematically, $H_0$ and $\Omega_m$ are either integration constants arising from solving ordinary differential equations or are directly related to integration constants. Seen in this context, if fits of the $\Lambda$CDM model to cosmological probes at different redshifts lead to different $(H_0, \Omega_m)$ parameters, this is a mismatch between mathematics and observation.
Here, in mock observational Hubble data (OHD) (geometric probes of expansion history) we demonstrate evolution in distributions of best fit parameters with effective redshift. As a result, considerably different $(H_0, \Omega_m)$ best fits from Planck-$\Lambda$CDM cannot be precluded in high redshift bins. We explore if OHD, Type Ia supernovae and standardisable quasar samples exhibit redshift evolution of best fit $\Lambda$CDM parameters. In all samples, we confirm a decreasing $H_0$ and increasing $\Omega_m$  trend with increasing bin redshift. Through comparison with mocks, we confirm that similar behaviour can arise randomly within the flat $\Lambda$CDM model with probabilities as low as $p = 0.0021$ ($3.1 \, \sigma$). We present complementary profile distribution analysis confirming the shifts in cosmological parameters in high redshift bins. In particular, we identify a redshift range where Planck $(H_0, \Omega_m)$ values are disfavoured at $99.6 \%$ ($2.9 \sigma$) confidence level in a combination of OHD and supernovae data.
\end{abstract}

\maketitle

\section{Introduction}

Cosmologists are currently debating  tensions within the flat $\Lambda$CDM cosmology; the two most serious concern the Hubble constant $H_0$ and the $S_8 := \sigma_8 \sqrt{\Omega_m/0.3}$ parameter \cite{DiValentino:2021izs, Abdalla:2022yfr}.\footnote{$S_8$ tension is less well established, see \cite{Nunes:2021ipq}.} These tensions have been framed as disagreements between the early (high redshift) and late (low redshift) Universe \cite{Verde:2019ivm}. In particular, local $H_0$ values \cite{Riess:2021jrx, Freedman:2021ahq, Pesce:2020xfe, Kourkchi:2020iyz, Blakeslee:2021rqi} are universally biased to larger values than Planck-$\Lambda$CDM \cite{Planck:2018vyg}. Observations at different redshifts have shown that $H_0$ evolves with effective (binned) redshift in the flat $\Lambda$CDM model \cite{Wong:2019kwg, Millon:2019slk, Krishnan:2020obg, Dainotti:2021pqg, Horstmann:2021jjg, Dainotti:2022bzg, Colgain:2022nlb, Wagner:2022etu, Jia:2022ycc} (see also \cite{Hu:2022kes}). If this trend is not due to observational selection biases, and it is intrinsic, this behaviour is indicative of model breakdown \cite{Krishnan:2020vaf, Krishnan:2022fzz}.

The flat $\Lambda$CDM model Hubble parameter $H(z)$ is specified by two constant fitting parameters $(H_0, \Omega_{m})$ or $(A,B)$,
\be\label{LCDM}
\begin{split}
H(z)^2 &= H_0^2 \left[1 - \Omega_{m} + \Omega_{m} (1+z)^3 \right], \\
   &= A +B(1+z)^3.  
\end{split}
\ee
The parameter $A:=H_0^2(1-\Omega_m)$ is attributed to dark energy (DE), while the matter sector $B:=H_0^2 \Omega_m$ scales as $(1+z)^3$ and $\Omega_{m}$ is bounded, $0 \leq \Omega_m \leq 1$. One can relax this constraint by allowing negative energy densities, but interpretation is problematic.\footnote{Later we show that mock realisations can easily violate this bound at higher redshifts. If the same trend is observed in observed data, does this immediately falsify flat $\Lambda$CDM?}
Observe that DE  becomes irrelevant at higher redshifts, where $A \ll B (1+z)^3$ for reasonable values of $\Omega_{m}$. On the other hand, note that at higher redshifts  $H(z)^2 \sim B (1+z)^3$, \footnote{{Throughout the text we use the symbol $\sim$ to highlight equivalences that are approximate.}} so the combination $\Omega_m h^2$, with $h:=H_0/100$, is the relevant quantity. Exploiting these facts, it was recently argued that increases in $\Omega_{m}$ (decreases in $H_0$) with effective redshift may be inherent to the flat $\Lambda$CDM model \cite{Colgain:2022nlb}. Here, we study
$\Lambda$CDM mocks binned by redshift to uncover the mathematical fact that the probability of Planck values $\Omega_m \sim 0.3$ decreases as we increase bin redshift. As a result, some evolution away from $\Omega_m \sim 0.3$ should be expected in best fits of purely high redshift observations.

Armed with this analytic insight, we turn to observed data in order to ascertain whether the same trend exists through comparison to mock simulations. We employ observational Hubble data (OHD), essentially cosmic chronometers \cite{Jimenez:2001gg} and baryon acoustic oscillations (BAO) \cite{Seo:2003pu,SDSS:2005xqv}, Type Ia supernovae (SNe) \cite{Pan-STARRS1:2017jku} and standardisable quasar (QSO) data sets \cite{Lusso:2020pdb}. Throughout we compare values of $(H_0, \Omega_{m})$ to mock simulations in the \textit{same} redshift range, where the base cosmology for the mock is fixed by the best fit parameters of the \textit{entire} data set. This allows us to confirm evolution between low and high redshifts in the sample. We provide complementary profile distribution analysis confirming the result.

Ultimately, while the fit of the overall sample to flat $\Lambda$CDM is largely dictated by the redshift range with greater density of data points, we will see that in sparser redshift ranges, the data prefers different cosmological parameters. In particular, we find probabilities as low as $p = 0.021$ (OHD), $p = 0.081$ (SNe) and $p = 0.019$ (QSOs), respectively, that mock data leads to similar values of $(H_0, \Omega_{m})$ as observed data. Combining the independent probabilities using Fisher's method, one arrives at the probability $p=0.0021$ ($3.1 \, \sigma$) that such an evolution indeed exists within flat $\Lambda$CDM. {In addition, we revisit the findings with profile likelihoods/distributions finding that Planck best fit values are disfavoured at $99.6\%$ ($2.9 \sigma$ for normal distributions) confidence level based on a combination of OHD and SNe data alone. This provides a sanity check using standard frequentist methodology.} An explanation in terms of selection biases is plausible for SNe, e. g.  \citep{Dainotti:2021pqg,Dainotti:2022bzg}, but similar effects must impact cosmic chronometers, BAO, etc. Our mock analysis shows that \textit{without selection biases}, evolution away from Planck values should be expected.

\section{Mock Data}
\label{sec:mock_data}
 Consider a simple data fitting exercise, where one takes Dark Energy Spectroscopic Instrument (DESI) forecasts for $H(z)$ errors $\sigma_{H(z_i)}$ at redshifts $z_i$ in the range $0.05 \leq z_i \leq 3.55$ \cite{DESI:2016fyo}. Next, adopt Planck values \cite{Planck:2018vyg}, $H_0 = 67.36$, $\Omega_m = 0.315$, for an underlying model and generate $H(z_i)$ values in a normal distribution about the Planck-$\Lambda$CDM model using the errors $\sigma_{H(z_i)}$ as the standard deviation at each $z_i$. Throughout we fix the parameters for the underlying cosmology and do not pick $(H_0, \Omega_m)$ in a distribution. Picking $(H_0, \Omega_m)$ in a distribution adds randomness, but this randomness is expected to be subleading to the randomness introduced in the shifts of the data points. For each realisation of mock data, separate the data into four bins, concretely $0 < z < 0.8$, $0.8 \leq z < 1.5$, $1.5 \leq z < 2.3$ and $2.3 \leq z < 3.6$. This ensures a similar number of data points in each bin. Finally, fit the parameters $(H_0, \Omega_m)$ from (\ref{LCDM}) to the data in each bin with a Gaussian prior on  $\Omega_{m} h^2 = 0.1430 \pm 0.0011$ \cite{Planck:2018vyg}. Note that the prior only provides guidance for the high redshift behaviour of $H(z)$ and its omission cannot change results (see appendix). Repeat the process a few thousand times and record the distribution of best fit values of $(H_0, \Omega_{m})$ for each bin.

\begin{figure}
   \centering
\begin{tabular}{c}
\includegraphics[width=70mm]{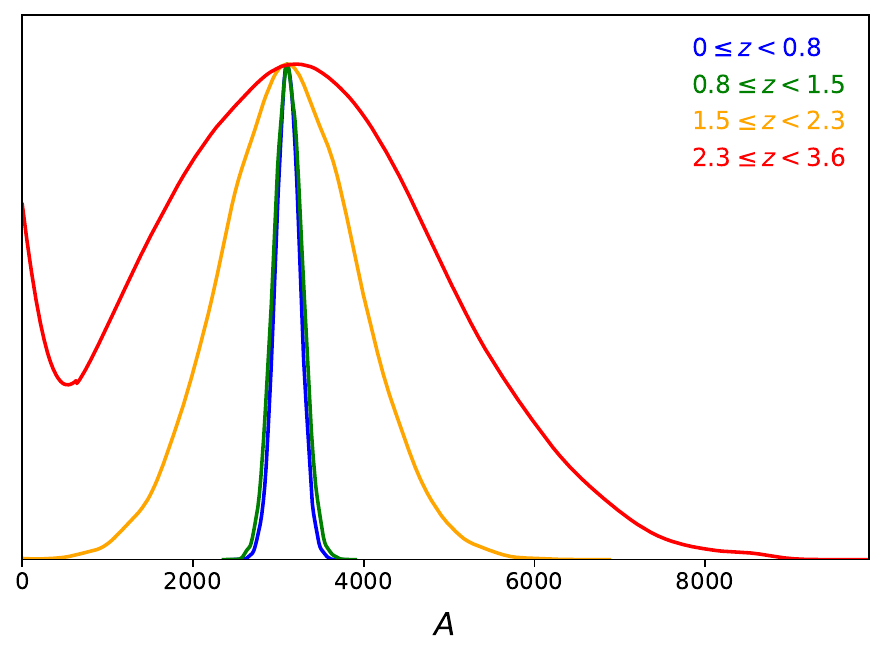} \\
\includegraphics[width=70mm]{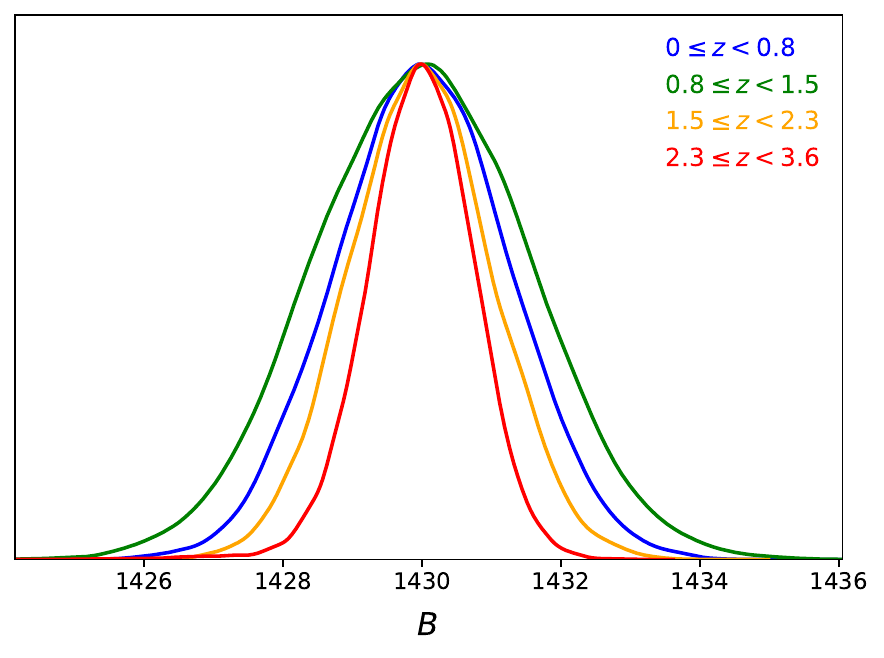}
\end{tabular}
\caption{Distributions of $A = H_0^2 (1- \Omega_{m}) $ and $B = H_0^2 \Omega_{m}$ parameters reconstructed from mock simulations of the Planck-flat-$\Lambda$CDM model in different redshift bins.}
\label{fig:AB_dist} 
\end{figure}

Before turning our attention to $H_0, \Omega_m$ best fit distributions, 
let us report on the (unnormalised) distributions for $A,B$. Fig. \ref{fig:AB_dist}, produced with \textit{GetDist} \cite{Lewis:2019xzd},  demonstrates that both $A$ and $B$ are Gaussian by inspection, except where $A$ is impacted by the boundary at $A=0$. Note, we have imposed a Gaussian prior on $B$, so $B$ being Gaussian is expected. Observe that the distributions in $A$ and $B$ spread and narrow, respectively, with increasing bin redshift. Interestingly, the distribution in $B$ spreads from bin 1 to bin 2 before narrowing in bins 3 and 4. This apparently contradicts our claim that $A$ spreads and $B$ narrows, but it can be traced to fractional error differences with redshift in the DESI forecast \cite{DESI:2016fyo}. If one ensures data with the \textit{same} fractional errors in all bins, then $A$ spreads and $B$ narrows with redshift. We demonstrate this in the appendix. This outcome is expected as the $\Lambda$CDM model (\ref{LCDM}) transitions from a two-parameter to an effective one-parameter model at high redshift. We have checked that $A$ and $B$ are uncorrelated (see appendix). We also see that $A$ grows a non-Gaussian tail around the $A=0$ ($\Omega_m = 1$) region at higher redshift bins. This comes about as a Gaussian with a wide spread probes the $A<0$ region with a growing probability in higher $z$ bins, which we have dubbed a `pile up' feature. Moreover,  the width of the Gaussian distribution for $B= H_0^2 \Omega_m$ reduces as we go to higher redshift bins and hence we know $B$ with a better precision in the higher redshift bins. Higher redshift spread in $A= H_0^2-B$ then yields spread in both $H_0$ and $\Omega_m$ values.

\begin{figure}
   \centering
\begin{tabular}{c}
\includegraphics[width=70mm]{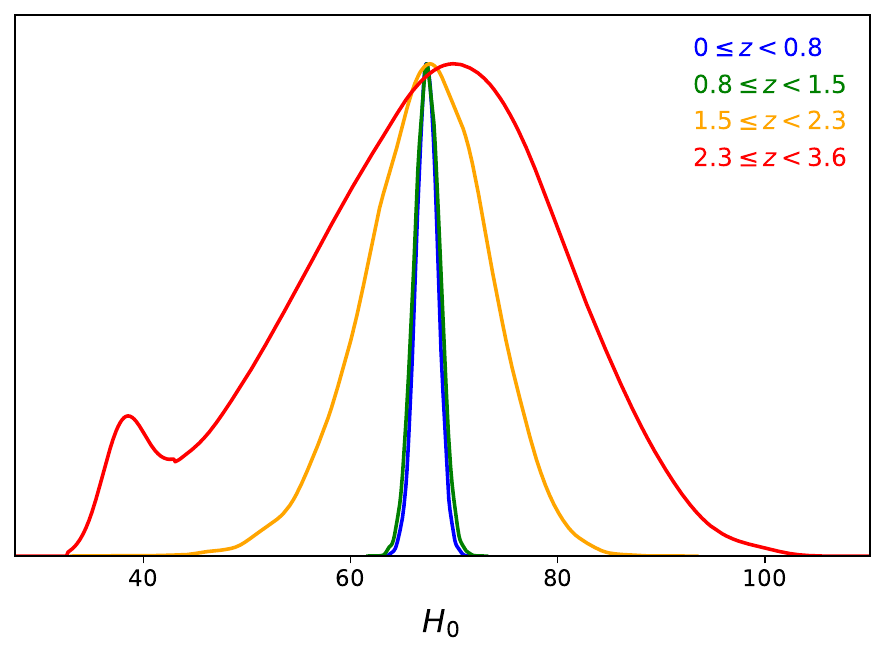} \\
\includegraphics[width=70mm]{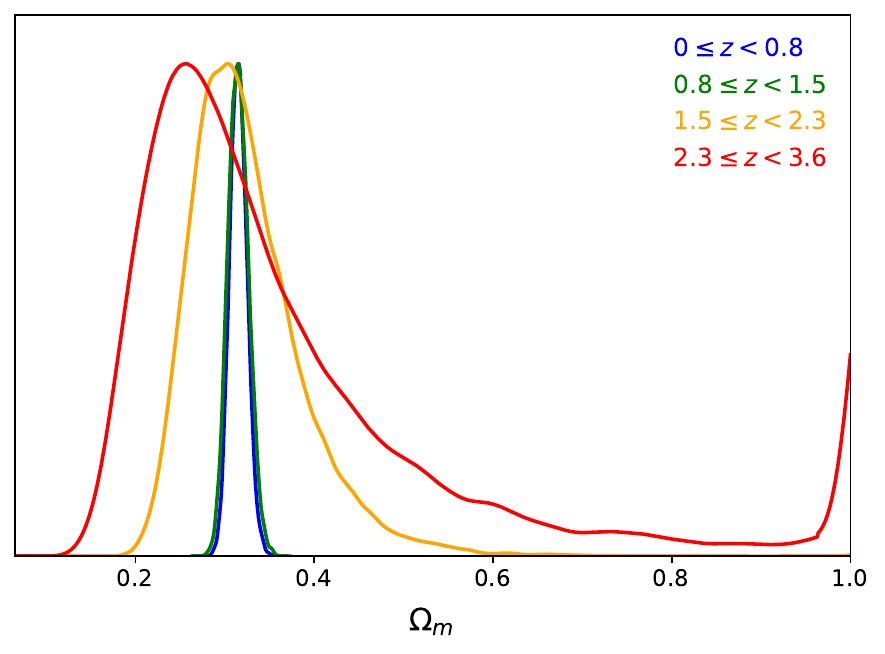}
\end{tabular}
\caption{Distributions of the cosmological parameters in different redshift bins. The `pile up' at $\Omega_{m} \sim 1$ and $H_0 \sim 37.8$ km/s/Mpc is due to $\Omega_m > 1$ best fits being restricted to the bound $\Omega_{m} = 1$.}
\label{fig:H0Om_dist} 
\end{figure}

In Fig. \ref{fig:H0Om_dist} we show the same distribution in $(H_0, \Omega_m)$ parameters. It is evident that both $H_0$ and $\Omega_{m}$ develop long non-Gaussian tails in the direction of smaller $H_0$ and larger $\Omega_{m}$ despite input Planck values in the mocking procedure, confirming our analytic expectations discussed above. This is easily explained. Since $\Omega_m h^2$ is well constrained, best fit $(H_0, \Omega_m)$ values inhabit a $\Omega_m H_0^2 \sim$ constant curve or banana. Nevertheless, as the banana stretches, configurations move from the peak to the extremities, leading to shifts in the peak when projected onto the $H_0$ and $\Omega_m$ axes. Thus, the $\Omega_m$ peak shifts to lower values, whereas the $H_0$ peak shifts to higher values. This comes from a ``projection effect" in the mock data. The pile up at $\Omega_m = 1$ is an artefact of our priors, but this can be relaxed without changing the conclusions. See \cite{Colgain:2022tql} for a more complete analysis. Our analysis here only concerns $H(z)$, but angular diameter distance $D_{A}(z)$ constraints, and the combination $H(z) + D_{A}(z)$, are studied in \cite{Colgain:2022tql}.

Since our mocking procedure is the same in each bin, while neither the number of data points nor the fractional errors change greatly (see \cite{DESI:2016fyo}), one concludes that the behaviour is generic to the flat $\Lambda$CDM model. {Note also that selection biases do not impact mocks.}
Moreover,  the same argument can be run for \textit{any} mock input parameters $(H_0, \Omega_m)$. The main message is that even in a Universe statistically consistent with Planck-$\Lambda$CDM by construction, unfamiliar best fit values can easily be returned in data fitting. Furthermore, best fits in the $\Omega_m > 1$ regime of parameter space are possible. See related analysis with Pantheon+ SNe \cite{Malekjani:2023dky}.

{To avoid confusion, the analysis in this section and its relation to the rest of the paper can be summarised as follows. Best fits of Planck-$\Lambda$CDM mocks in high redshift bins generically lead to non-Gaussian $H_0$ and $\Omega_m$ distributions if one works with either OHD or luminosity/angular diameter distance constraints. Moreover, even in mocks one can find unexpectedly small and large best fit values of $H_0$ and $\Omega_m$, respectively. In the next section we study best fits of observed data and quantify the unlikeliness of the best fits against mocks in the same redshift range with the same fractional errors for the data. This section basically explains why the distributions in Figs. \ref{fig:OHDsim}, \ref{fig:SNsim} and \ref{fig:QSOsim} are non-Gaussian. This is largely a technical point that one does not need to process to digest later results. Irrespective of the shape of the distributions, the $p$-values in the next section are based on an ordering of best fit values from mocks, but the shape of the distribution is a secondary concern, since one can define percentiles without reconstructing a probability density function (PDF).}

\section{Observed Data}
\label{sec:data}
Having uncovered a general feature for $H(z)$ constraints confronted to the flat $\Lambda$CDM model, {i. e. an increase in the likeliness of smaller $H_0$ and larger $\Omega_m$ values in high redshift bins, we now explore the extent to which this feature is manifest in observed OHD.} In \cite{Colgain:2022tql} we show that $D_{L}(z) \propto D_{A}(z)$ constraints confronted to flat $\Lambda$CDM exhibit similar features, which justifies studying Type Ia SNe and QSOs. 

\subsection{Comments on Methodology}
When one finds an anomaly in cosmological data, for example CMB anomalies \cite{Eriksen:2003db}, one typically resorts to mock simulations to assign a statistical significance to the feature. Here our focus will be a decreasing $H_0$/increasing $\Omega_m$ trend in best fits with increasing effective redshift. Moreover,
as we have seen, one encounters non-Gaussian distributions in exclusively high redshift bins (see also \cite{Colgain:2022tql}). As a result, while best fits, i. e. the extrema of $\chi^2$, are expected to be robust within machine precision,\footnote{One can test this by initialising the $\chi^2$-minimsation algorithm from different points in parameter space and checking that one recovers best fit parameters that are close in value. See analysis in \cite{Malekjani:2023dky}.} estimating errors as is usually done in cosmology is difficult. More explicitly, Fisher matrix leads to unrepresentative Gaussian errors, while Markov Chain Monte Carlo (MCMC) inferences are prone to degeneracies/projection effects that distort inferences. Moreover, with broad distributions it is possible that MCMC inferences are simply tracking the priors (e. g. see Fig. 2 of \cite{Malekjani:2023dky}) and the peaks of distributions are not guaranteed to coincide with the minimum of the $\chi^2$ \cite{Colgain:2023bge}. We highlight an explicit difficulty with MCMC analysis in the appendix. 

Given the difficulties with conventional techniques, here we resort to mocks that allow us to generate a large number of best fits that are \textit{statistically consistent (by construction) with no evolution of cosmological parameters}. We make direct comparison between best fits from mocks and observed data \textit{in the same redshift range with the same data points and errors}. This allows us to rank mock best fits of $H_0$ and $\Omega_m$ in descending and ascending order, respectively, and identify the percentile where observed data best fits appear. This gives us a probability for finding similar best fits \textit{assuming no evolution in the sample}. Note, just as the shape of a PDF of heights of children in a class is irrelevant in such an exercise, the same logic also applies here. {Bluntly put, elementary school teachers can assign a percentile to the height of a student without necessarily understanding the concept of a PDF.}

In all samples, we note that the probabilities (see Tables \ref{tab:OHD}, \ref{tab:SN} and \ref{tab:QSO}) of finding observed data best fits as extreme in mock data decrease as the effective redshift of subsamples becomes less representative of the full sample. This is expected if the trend is due to shifts in best fit cosmological parameters. However, the probabilities do not decrease indefinitely, and our results show that the probabilities increase again in the smallest subsamples, which we attribute to noise. It is intuitive that any signal in the data eventually disappears due to statistical fluctuations in small samples. Furthermore, given the probabilities decrease with increasing difference in effective redshift between subsample and the full sample, this means that this probability is bounded below. Thus, we do not pick redshift ranges by hand, but they emerge from the data as the redshift ranges where best fits in a subsample are least representative of the full sample. In other words, one can give a lower bound on probabilities and this lower bound is expected to be well defined. One may worry that the $p$-values we record are artefacts of degeneracies between $H_0$ and $\Omega_m$ that drive best fits in mock simulations along unconstrained directions in parameter space. To negate this concern in section \ref{sec:PD} we revisit the statistical significance of the best fits using profile distributions. In contrast to mock simulations, the later analysis only makes use of a single realisation of (observed) data.

We impose a strong Planck $\Omega_m h^2= 0.1430 \pm 0.0011$ prior, which constrains best fits to a curve in the $(H_0, \Omega_m)$-plane. In tandem we start $\chi^2$-minimisation for each realisation of the data, either observed or mock, from the best fits of the full sample. As a result, if there is little or no evolution, one expects the best fits to not move far from the initial guess. In other words, we bias the initial guess towards no evolution. However, our strong Planck $\Omega_m h^2$ prior reduces the fitting procedure to an effective 1-dimensional fit in the $(H_0, \Omega_m)$-plane. What this means in practice is that we may find false minima, but these minima are the closest to the input parameters. Nevertheless, we think false minima are unlikely, given the effective 1-dimensional nature of the fitting.  More concretely, note that even in OHD, we are performing an effective 1-dimensional fit with no less than 6 data points and it is hard to imagine that the outcome is not unique modulo machine precision. Indeed, we will confirm later with profile distributions that we recover the best fits from the MCMC chain, which rules out false minima.

{Before proceeding, we make some explicit comments on the mocking procedure initially introduced in section \ref{sec:mock_data}. For each sample, we fit the full sample to identify best fit parameters for the $\Lambda$CDM model. The exception here is the QSO sample where we consider a redshift range up to the point where we find an $\Omega_m = 1$ best fit, and do not consider the full data set. Note, the best fits should be representative values if $(H_0, \Omega_m)$ do not change with effective redshift through the samples. Then, for all the data points in the redshift ranges of interest in Tables \ref{tab:OHD}, \ref{tab:SN} and \ref{tab:QSO}, we randomly generate new data points in a normal distribution about the best fit $\Lambda$CDM model using the cropped covariance matrix, a process we repeat thousands of times to build up the histograms in the samples presented in Figs. \ref{fig:OHDsim}, \ref{fig:SNsim} and \ref{fig:QSOsim}. For Pantheon SNe, a covariance matrix is available. For OHD, the covariance matrix is diagonal, so mirroring earlier analysis in section \ref{sec:mock_data}, we generate new data points in a normal distribution about the best fit cosmological model where the standard deviation coincides with the errors. For QSOs, there is a slight tweak to the mocking procedure, but we discuss it later.}  

\subsection{OHD} 
\label{sec:OHD}
Here, we make use of cosmic chronometer \cite{Stern:2009ep, Moresco:2012jh, Zhang:2012mp, Moresco:2016mzx, Ratsimbazafy:2017vga, Borghi:2021rft, Jiao:2022aep} and BAO data \cite{Gaztanaga:2008xz, Oka:2013cba, BOSS:2016zkm, Chuang:2012qt, BOSS:2016wmc, Blake:2012pj, Anderson:2013oza, Bautista:2017zgn, BOSS:2014hwf, BOSS:2013igd}. More precisely, we work with the $H(z)$ BAO determinations compiled in Table 2 of \cite{Magana:2017nfs}, where observations have been homogenised to be consistent with a uniform Planck inference of the sound horizon \cite{Planck:2015fie}. We added the newer constraint from eBOSS Quasar \cite{Hou:2020rse, Neveux:2020voa}, which we appropriately adjusted for the sound horizon, $H(z=1.48) = 153.59 \pm 8.27$. {In addition to 21 BAO data points, we make use of 33 cosmic chronometer data points. Concretely, we utilise the 35 data points in Table 1.1 of \cite{Moresco:2023zys}, where we omit two of the most recent additions at $z = 0.75$ and $z = 1.26$. As is clear from Table 1.1, there is still overlap in the remaining data points at $z=0.75$ and $z = 0.8$, but this does not affect the high redshifts where we see departures from Planck behaviour.} Our total sample has 54 OHD sources. Moreover, we have checked that replacing earlier Lyman-$\alpha$ BAO \cite{Bautista:2017zgn, BOSS:2014hwf, BOSS:2013igd} with the latest constraints \cite{duMasdesBourboux:2020pck} does not {greatly} change the results,\footnote{{Replacing historical Lyman $\alpha$-BAO with later constraints shifts the best fit value of $\Omega_m$ to lower values, but a $>2 \sigma$ discrepancy with Planck is still evident for OHD with $z > 1.45$ \cite{Colgain:2023bge}.}} so we work with the earlier determinations collated in  \cite{Magana:2017nfs}.  

\begin{figure}
   \centering
\includegraphics[width=90mm]{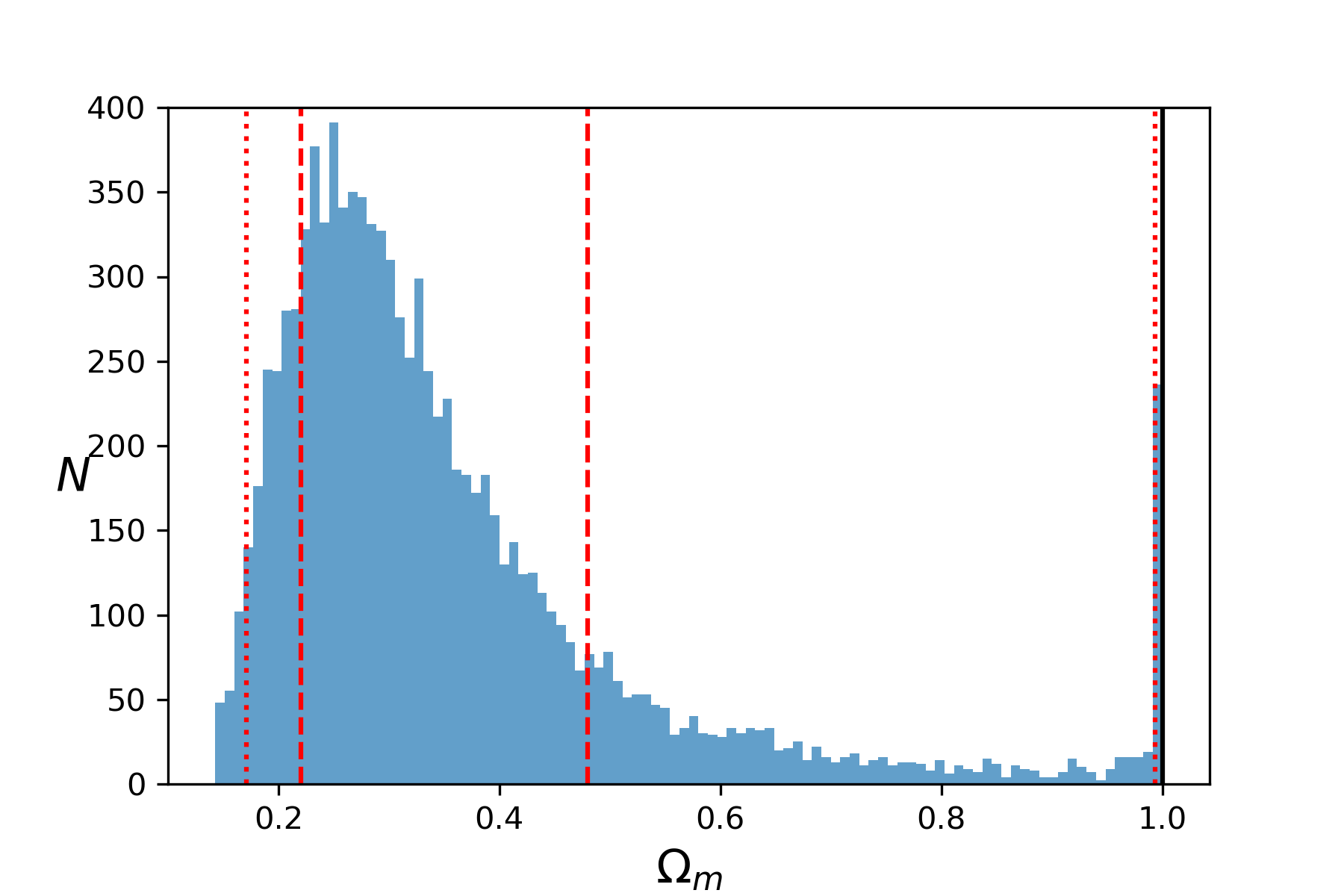}
\caption{Comparing 10,000 mock simulations with the best fit value of $\Omega_{m}$ from OHD data (black line) for the bin $1.45 \leq z \leq 2.36$. Dashed and dotted lines denote the $(2.3, 15.9, 84.1, 97.7)$ percentiles.}
\label{fig:OHDsim} 
\end{figure}

\begin{figure}
   \centering
\includegraphics[width=90mm]{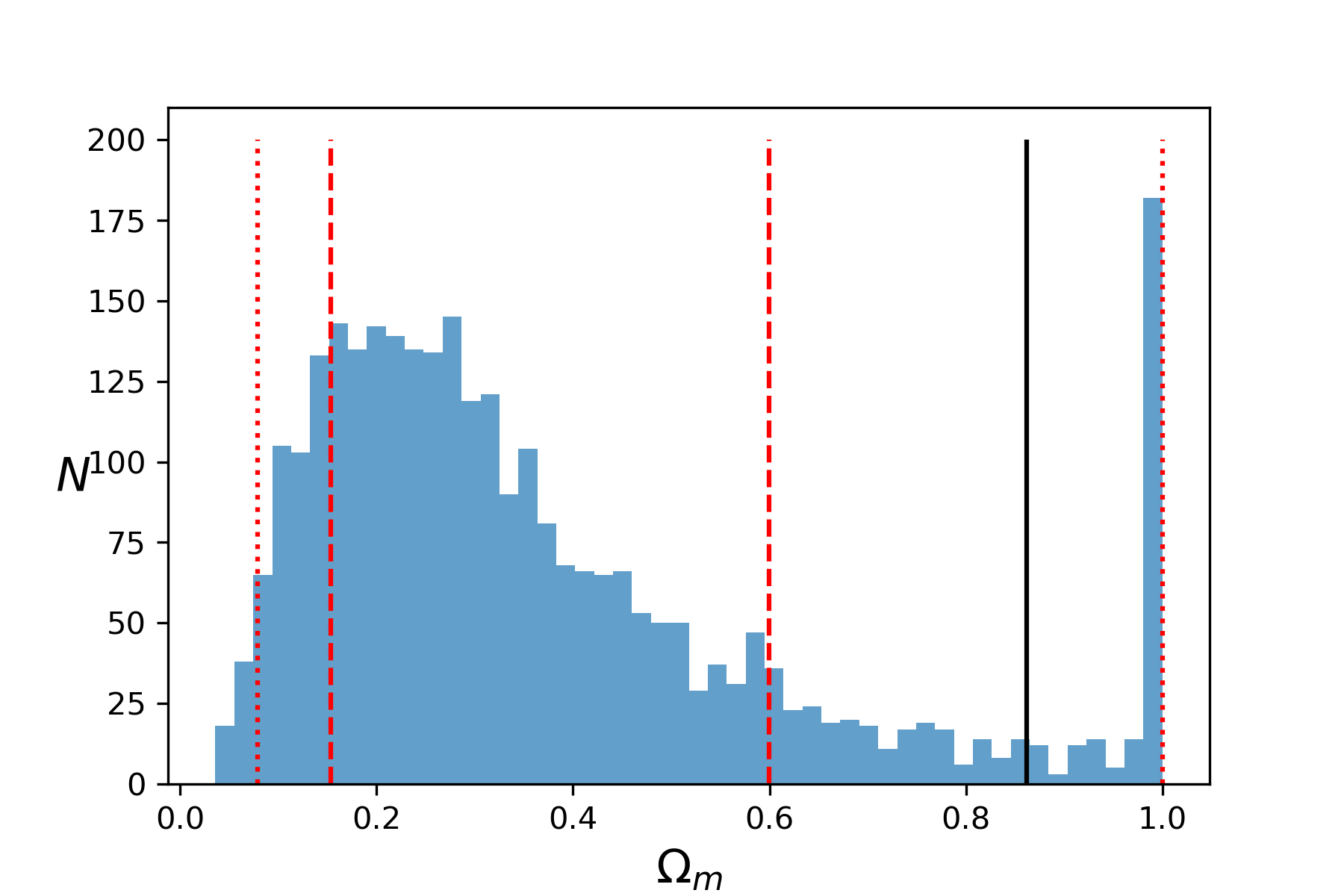}
\caption{Comparing 3,000 mock simulations with the best fit value of $\Omega_{m}$ from SNe data (black line) for the bin $0.95 < z \leq 2.26$. Dashed and dotted lines denote the $(2.3, 15.9, 84.1, 97.7)$ percentiles corresponding to $1 \sigma$ and $2  \sigma$ confidence intervals for a Gaussian distribution.}
\label{fig:SNsim} 
\end{figure}

First, we identify the best fit values of the cosmological parameters for the full sample, $(H_0, \Omega_{m}) = (69.11, 0.299)$, where it is worth noting that $\Omega_{m} h^2 = 0.1428$, consistent with the prior. Next, we repeat the mocking and binning procedure outlined earlier with the new input parameters $(H_0, \Omega_m) = (69.11, 0.299)$. Following  \cite{Colgain:2022nlb} we impose a low redshift cutoff to remove sources below a given $z$ and isolate high redshift bins. In each bin we compare the best fit values from the real data and flat $\Lambda$CDM mocks \textit{in the same bin with the same number of data points and same errors} in order to establish the probability of recovering \textit{the same or larger} $\Omega_m$ and \textit{the same or smaller} $H_0$ values. In the event of saturation of the bound $\Omega_m = 1$, this means that our probabilities are over-estimated, i. e. too large, since allowing $\Omega_m > 1$ permits further ordering of the values piled up at $\Omega_m = 1$. The results are shown in Table \ref{tab:OHD}, where it is clear that $(H_0, \Omega_m)$ best fits are evolving in the real data. For easy comparison throughout, we include the best fits for the full sample in tables, but do not assign any probabilities. Understandably, the probability of recovering similar values from mocks decreases with redshift up to a point where statistical fluctuations dominate and the probability increases again. Fig. \ref{fig:OHDsim} provides visual confirmation that despite the long tails, a bin exists where the real values are unexpected at {95\% confidence level ($\gtrsim 2 \sigma$ for a normal distribution \footnote{{As explained in section \ref{sec:mock_data}, none these distributions are expected to  be Gaussian.}})}. This points to redshift evolution in the sample.

\begin{table}
    \centering
    \begin{tabular}{c|ccc}
         $z$ & $H_0$ (km/s/Mpc) & $\Omega_m$  & Probability \\
         \hline 
         $0 \leq z \leq 2.36$ $(54)$ & $69.11$ &  $0.299$ & $-$ \\
         $0.5 \leq z \leq 2.36$ $(28)$ & $69.68$ &  $0.294$ & $0.646$ \\
          $0.7 \leq z \leq 2.36$  $(18)$ & $65.67$ &  $0.331$ & $0.326$ \\
           $1 \leq z \leq 2.36$ $(11)$ & $61.27$ &  $0.380$ & $0.258$ \\
           $1.2 \leq z \leq 2.36$ $(10)$ & $53.91$ &  $0.491$ & $0.120$ \\
            $1.4 \leq z \leq 2.36$ $(8)$ & $41.55$ &  $0.828$ & $0.037$ \\
            $1.45 \leq z \leq 2.36$ $(7)$ & $37.80$ &  $1$ & $0.021$ \\
             $1.5 \leq z \leq 2.36$ $(6)$ & $37.80$ &  $1$ & $0.069$ \\
    \end{tabular}
    \caption{Best fit cosmological parameters for different redshift ranges of OHD. Throughout, we impose the Planck prior, $\Omega_{m} h^2 = 0.1430 \pm 0.0011$. Flat $\Lambda$CDM simulations based on best fit parameters over the entire redshift range, $0 < z \leq 2.33$, allow us to establish the probability of higher $\Omega_m$ and lower $H_0$ values in real data. The OHD count in each bin is denoted in brackets.}
    \label{tab:OHD}
\end{table}

\subsection{Type Ia SNe} 
We revisit the analysis of the Pantheon data set \cite{Pan-STARRS1:2017jku} with $1048$ SNe conducted in \cite{Colgain:2022nlb} (see also \cite{Dainotti:2021pqg, Horstmann:2021jjg}) in order to introduce a high redshift Planck prior on $\Omega_{m} h^2$ \cite{Planck:2018vyg}. Note, to do so, we treat the absolute magnitude of Type Ia SNe $M_{B}$ as a nuisance parameter. This gives SNe data the freedom to adjust $H_0$ so that the high redshift behaviour is always the same as Planck, otherwise the analysis is the same as before. Alternatively put, we have an additional nuisance parameter, but its role is simply to adopt the value that best accommodates fits in the $(H_0, \Omega_m)$-plane, where we are still confronted with an effective 1-dimensional fit. We identify the best fit parameters $(H_0, \Omega_m, M_B) = (69.26, 0.298, -19.37)$, construct mock realisations in bins, which one compares to the real values. Throughout we allow for statistical and systematic uncertainties by cropping the Pantheon covariance matrix accordingly to fit the redshift bin. The results are shown in Table \ref{tab:SN} and Fig. \ref{fig:SNsim}, where the same trend as the OHD data is evident.

\subsection{Standardisable QSOs} 
Finally we turn our attention to QSOs standardised through the Risaliti-Lusso proposal \cite{Risaliti:2015zla, Risaliti:2018reu}. We refer readers to the original texts for methodology. Objectively, QSOs constitute emerging cosmological probes \cite{Moresco:2022phi,  Dainotti:2022wli} and are understandably less well developed than the SNe and BAO; nevertheless, even now SNe remain a work in progress \cite{DES:2022tgg}. In particular, there is considerable intrinsic scatter in the QSO data and there is an ongoing debate about the standardisability of the Risaliti-Lusso QSOs \cite{Khadka:2020vlh, Khadka:2021xcc, Khadka:2020tlm, Dainotti:2022wli, Petrosian:2022tlp}. In contrast to OHD and SNe, which have lower error-weighted (effective) redshifts of $z_{\textrm{eff}} \sim 0.5$ and $z_{\textrm{eff}} \sim 0.3$, respectively, the QSO  sample \cite{Lusso:2020pdb} is larger (2421 sources) and has a higher effective redshift $z_{\textrm{eff}} \sim 1.4$.  {The sample is too large to present in a table, but can be downloaded from the original source \cite{Lusso:2020pdb}.} Moreover, it is well documented that $\Omega_{m}$ adopts larger values than expected at higher redshifts \cite{Risaliti:2018reu, Yang:2019vgk} and that evolution happens within the QSO sample \cite{Risaliti:2018reu, Colgain:2022nlb}. The key point here is that any evolution of $\Omega_m$ with effective redshift may be telling us less about QSOs and more about the flat $\Lambda$CDM model.  

\begin{table}
    \centering
    \begin{tabular}{c|ccc}
         $z$ & $H_0$ (km/s/Mpc) & $\quad\Omega_{m}$ & Probability  \\
         \hline
         $0 < z \leq 2.26$ $(1048)$ & $69.26$ & $\quad 0.298$ & $-$ \\
         $0.7 < z \leq 2.26$ $(124)$ & $64.37$ & $\quad 0.345$ & $0.381$ \\
         $0.8 < z \leq 2.26$ $(82)$ & $58.99$ & $\quad 0.411$ & $0.258$ \\
         $0.9 < z \leq 2.26$ $(49)$ & $45.88$  & $\quad 0.679$ & $0.117$ \\
         $0.95 < z \leq 2.26$ $(34)$ & $40.73$ & $\quad 0.862$ & $0.081$ \\
         $1 < z \leq 2.26$ $(23)$ & $43.16$ & $\quad 0.768$ & $0.170$
    \end{tabular}
    \caption{
    Same as Table \ref{tab:OHD} but for Pantheon SNe. We treat the absolute magnitude $M_{B}$ as an additional nuisance parameter when we fit mock realisations and real data. We quote the probability of larger values of $\Omega_{m}$ and lower values of $H_0$. SNe count is denoted in brackets.
    }
    \label{tab:SN}
\end{table}

\begin{figure}
   \centering
\includegraphics[width=90mm]{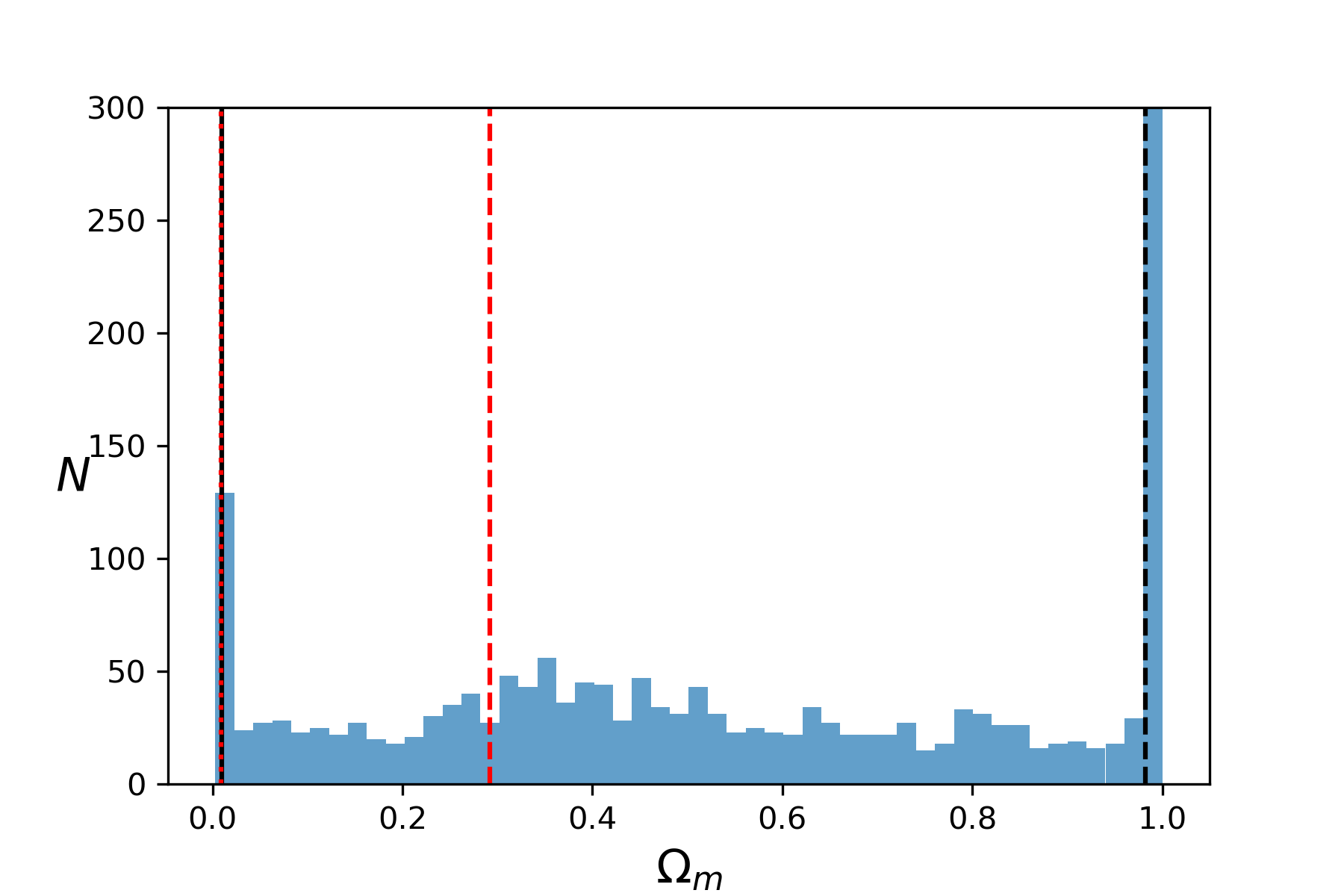}
\caption{A comparison between 3,000 mock simulations and the best fit value of $\Omega_{m}$ from QSO data (black line) for the bin $0 < z \leq 0.55$. Dashed and dotted red lines denote the $(2.3, 15.9)$ percentiles corresponding to $1 \, \sigma$ and $2 \, \sigma$ confidence intervals for a Gaussian distribution. The dashed black line denotes the median, $\Omega_m = 0.982$, which, as expected, is close to the mock input $\Omega_m = 1$. In contrast to Fig. \ref{fig:OHDsim} and Fig. \ref{fig:SNsim}, the PDF is flat (neglecting the impact of bounds), thus implying that $\Omega_m$ errors are large, i. e. $\Omega_m$ is poorly constrained, in the redshift range.}
\label{fig:QSOsim} 
\end{figure}

\begin{table}
    \centering
    \begin{tabular}{c|cccc}
         $z$ & $H_0$ (km/s/Mpc) & $\quad\Omega_{m}$ & Probability  \\
         \hline 
         $0 < z \leq 0.3$ $(56)$ & $406.41$ & $\quad 0.009$ & $0.073$ \\
         $0 < z \leq 0.5$ $(177)$ & $353.47$ & $\quad 0.011$ & $0.028$ \\
         $0 < z \leq 0.55$ $(233)$ & $433.91$ & $\quad 0.008$ & $0.019$ \\
         $0 < z \leq 0.6$ $(279)$ & $381.50$ & $\quad 0.010$ & $0.020$ \\
         $0 < z \leq 0.7$ $(398)$ & $73.40$ & $\quad 0.265$ & $0.096$ \\
         $0 < z \leq 0.8$ $(543)$ & $58.48$ & $\quad 0.418$ & $0.117$ \\
         $0 < z \leq 1$ $(826)$ & $40.69$ & $\quad 0.864$ & $0.400$ \\
          $0 < z \leq 1.4$ $(1326)$ & $37.82$ & $\quad 1.000$ & $-$ \\
    \end{tabular}
    \caption{
    Same as Table \ref{tab:OHD} but for Risaliti-Lusso QSOs. We treat $\beta$, $\gamma$ and $\delta$ (see \cite{Risaliti:2015zla} for definitions) as additional nuisance parameters when we fit mock realisations and real data. We quote the probability of \textit{lower} values of $\Omega_{m}$ and \textit{higher} values of $H_0$. QSO count is denoted in brackets.}
    \label{tab:QSO}
\end{table}

Our analysis here follows the earlier sections, but there is a key difference. Risaliti-Lusso QSOs return best fits of $\Omega_m \sim 1$ across the full sample \cite{Yang:2019vgk, Khadka:2020vlh, Khadka:2021xcc, Khadka:2020tlm}, whereas at lower redshifts $0 < z \lesssim 0.7$, one recovers Planck values, $\Omega_m \sim 0.3$ \cite{Colgain:2022nlb}; 
in accord with our earlier discussions and analyses. Thus, we start from the redshift range $0 < z \leq 1.4$ (1326 QSOs), where $\Omega_m$ hits the bound $\Omega_m = 1$, and identify the best fit parameters that serve as inputs for mocks, $(H_0, \Omega_{m}, \beta, \gamma, \delta) = (37.82, 1, 8.64, 0.61, 0.24)$. As before, $\beta$ is a nuisance parameter degenerate with $H_0$ (the analogue of $M_B$ in SNe), so once again the fit in the $(H_0, \Omega_m)$-plane is effectively 1-dimensional. To construct the mocks, we generate new UV fluxes $F_{UV}$ by picking values in a normal distribution about the original values with a standard deviation set to the error. Next, we generate corresponding central values for the X-ray fluxes $F_{X}$ through the relation \cite{Risaliti:2015zla, Risaliti:2018reu},  
\begin{equation}
    \log_{10} F_{X} = \beta  + \gamma \log_{10} F_{UV} +  (\gamma-1)  \log_{10} (4\pi D_{L}^2), 
\end{equation}
where $D_{L}(z)$ is the luminosity distance, before displacing the values with the standard deviation $\sqrt{\delta^2 + \sigma_{i}^2}$, where $\sigma_i$ is the error on $\log_{10} F_{X, i}$ at redshift $z_i$. 

In Table \ref{tab:QSO} we show the increasing (decreasing) trend of $\Omega_m$ ($H_0$) with effective redshift. Unexpectedly large values of $H_0$ and small values of $\Omega_m$ are driven partially by large intrinsic scatter in the QSO data and the Planck prior on $\Omega_m h^2$. Nevertheless, the trend in central values is the same and one notes that the probability of recovering the best fit values for real data decreases as the effective redshift of the bin decreases, confirming that the best fit values of the entire data set are less representative. In Fig. \ref{fig:QSOsim} we provide visual confirmation of this result in a given range, where it is notable that the $\Omega_m$ distribution is uniform between the bounds, thus underscoring how poorly QSO data constrains $\Omega_m$ in the corresponding redshift range. This is presumably due to the large scatter and fewer QSOs at lower redshifts. 

\section{Profile Distributions}
\label{sec:PD}
In this section we revisit earlier analysis from the perspective of profile distributions. The objective is to provide an alternative view on our mock simulation analysis where one may be worried that the low $p$-values are driven by noise in the mocks and the degeneracy or anti-correlation between $H_0$ and $\Omega_m$. Our methodology follows \cite{Gomez-Valent:2022hkb, Colgain:2023bge}. {The analysis is standard frequentist analysis (see section 4 of \cite{Herold:2021ksg}), but there is a small tweak. Instead of optimising, we bin the MCMC chain to construct the profiles, thereby ensuring as close a comparison as possible between Bayesian and frequentist methods; the MCMC chain is the input in both analyses (see \cite{Gomez-Valent:2022hkb} for further discussion).} The basic idea is to study the probability distribution
\begin{equation}
    \mathcal{P}(H_0, \Omega_m, \theta_i) = \exp \left( - \frac{1}{2} \chi^2 (H_0, \Omega_m, \theta_i) \right)
\end{equation}
where $\chi^2(H_0, \Omega_m, \theta_i)$ is the $\chi^2$ likelihood, which may depend on additional nuisance parameters, $\theta_i, i = 1, 2, \dots$, e.g the absolute magnitude $M_B$ from Type Ia SNe, {or $\beta, \gamma, \delta$ from standardisable QSOs}. The maximum value of $\mathcal{P}$ occurs at the $\chi^2$ minimum, $\mathcal{P}_{\textrm{max}} = e^{-\frac{1}{2} \chi^2_{\textrm{min}}}$. In contrast to the previous sections where $\chi^2_{\textrm{min}}$ is determined through gradient descent (optimisation), here we directly evaluate the $\chi^2$ likelihood on the MCMC chain to identify the minimum. We stress that this involves no optimisation, but if all analysis is consistent, we expect to recover profile distribution peaks that agree with best fits from Tables \ref{tab:OHD}-\ref{tab:QSO}. 

Our next step is to pick a parameter, identify its range, i. e. minimum and maximum value from the MCMC chain, and then divide the range of the parameter into approximately 200 bins centered on the parameter value at the centre of the bin. By increasing the length of the MCMC chain one can easily increase the number of bins. Focusing on $H_0$, we define the profile distribution for $H_0$ as
\begin{equation}
\tilde{\mathcal{P}}(H_0) = \exp \left( - \frac{1}{2} \chi_{\textrm{min}}^2 (H_0) \right)
\end{equation}
where $\chi^2_{\textrm{min}}(H_0)$ denotes the minimum value of $\chi^2$ along other directions in parameter space for the $H_0$ values in a given bin. If the bin is empty, as can happen frequently in the tails of distributions, we simply omit the bin. This accounts for missing dots in the later plots. At this stage, we can define the ratio 
\begin{equation}
    R(H_0) = \frac{\tilde{\mathcal{P}}(H_0)}{\mathcal{P}_{\textrm{max}}} = \exp \left(-\frac{1}{2} ( \chi^2_{\textrm{min}}(H_0) - \chi^2_{\textrm{min}}) \right).  
\end{equation}
We emphasise again that $\chi^2_{\textrm{min}}$ is the absolute minimum from the MCMC chain, whereas $\chi^2_{\textrm{min}}(H_0)$ is the minimum in a bin centered on $H_0$. The distribution $R(H_0)$ is peaked at $R(H_0) = 1$ by construction, since the absolute minimum of the $\chi^2$ must appear in one of the $H_0$ bins. What remains is to normalise $R(H_0)$ and turn it into a PDF:
\begin{equation}
w(H_0) = \frac{R(H_0)}{\int R(H_0) \, \textrm{d} H_0}, 
\end{equation}
where in the denominator the integral is over the full range of $H_0$ values from the MCMC chain. Bearing in mind that we have discretised the $H_0$ range in bins, this integral is most easily evaluated using Simpson's rule for numerical integration. One could worry about this approximation, but we will now integrate over $w(H_0)$, so that integrals in the numerators and denominator are performed to the same accuracy. Finally, to identify $68 \%, 95 \%$ and $99.7 \%$ confidence intervals ($1 \sigma, 2 \sigma$ and $3 \sigma$ for Gaussian distributions) for $H_0$, we identify $H_0^{(1)}$ and $H_0^{(2)}$ so that 
\begin{equation}
\begin{aligned}
    &\int_{H_0^{(1)}}^{H_0^{(2)}} w (H_0) \, \textrm{d} H_0 = I,  \quad w(H_0^{(1)}) = w(H_0^{(2)}),  \cr 
    &I \in \{ 0.68, 0.95, 0.997\}. 
\end{aligned}
\end{equation}
Expressions for $\Omega_m$ are defined in an analogous fashion.

Our treatment in this section will not be exhaustive and we focus largely on OHD and SNe data, since these data sets are most familiar to cosmologists. In short, our objective is to recover the high redshift best fits from Table \ref{tab:OHD} and \ref{tab:SN} and estimate the significance of the discrepancies directly from the profile distribution. It should be stressed that this exercise has been repeated with later and better quality data sets, where it was shown that mock simulations and profile distributions show good agreement \cite{Malekjani:2023dky,Colgain:2023bge}. {In addition, we revisit Tables \ref{tab:OHD}, \ref{tab:SN} and \ref{tab:QSO} to determine the $68 \%$ confidence intervals, which were omitted in the earlier tables. This provides confirmation that the best fits are discrepant with Planck outside of the errors at higher redshifts in OHD, SNe and QSOs. Given the different systematics across these observables, this cannot be a coincidence.}

\begin{figure}
   \centering
\begin{tabular}{c}
\includegraphics[width=70mm]{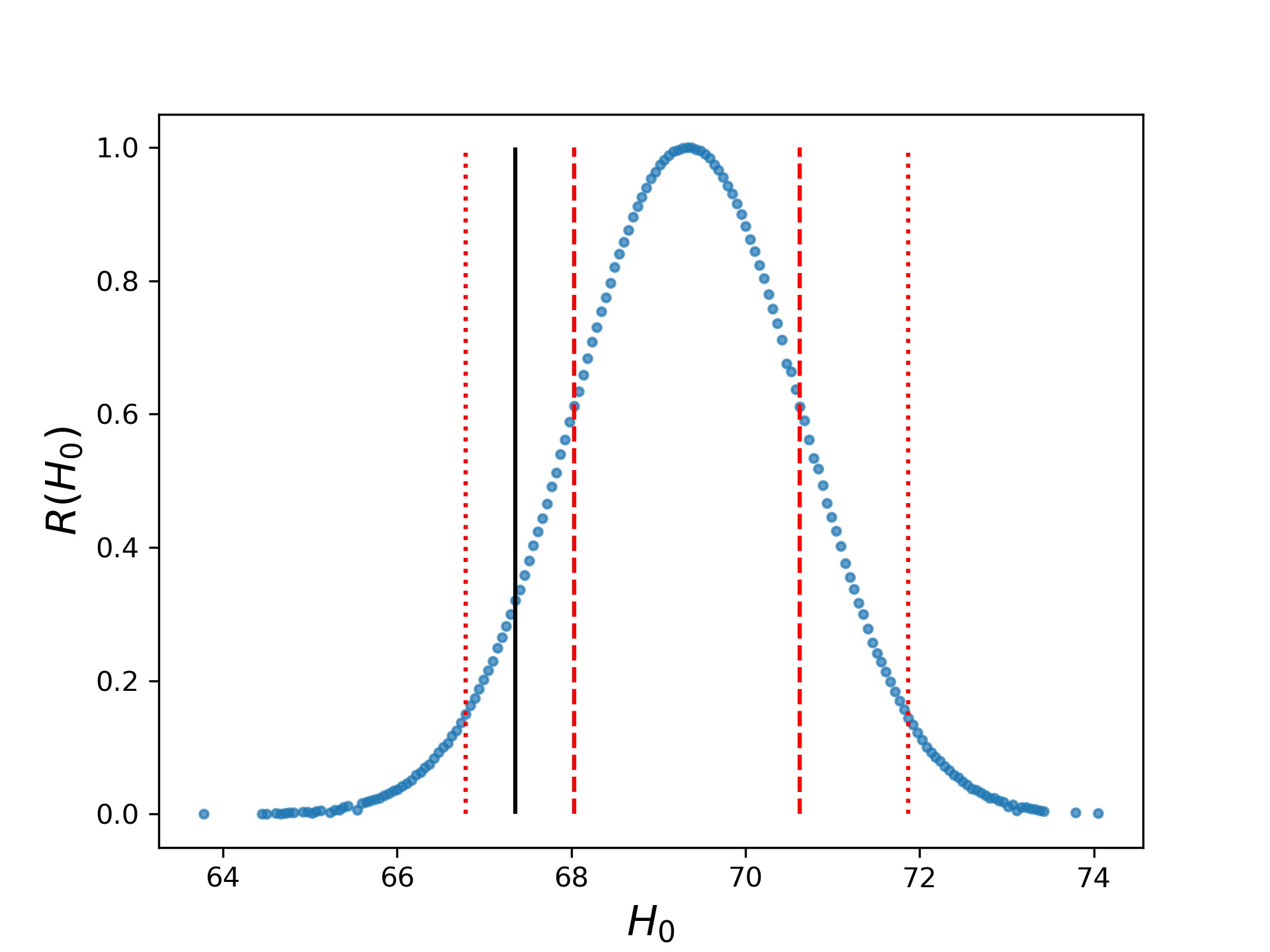} \\
\includegraphics[width=70mm]{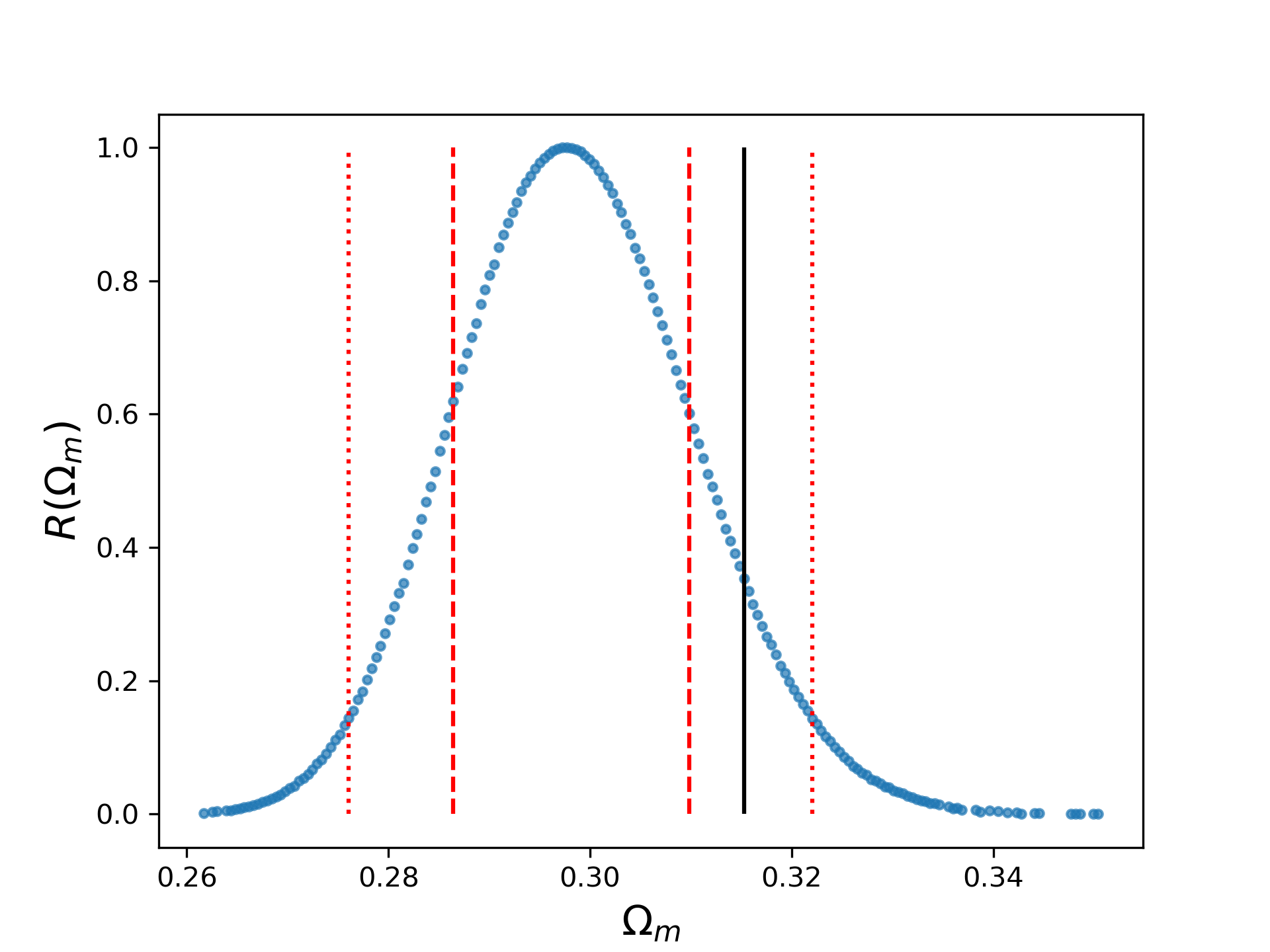}
\end{tabular}
\caption{$R(H_0)$ and $R(\Omega_m)$ distributions for OHD data with $z \leq 1.45$. Dashed and dotted red lines denote $68 \%$ ($1 \sigma$) and $95 \%$ ($2 \sigma$) confidence intervals. Black lines denote Planck best fit values.}
\label{fig:R_OHD_low} 
\end{figure}

\begin{figure}
   \centering
\begin{tabular}{c}
\includegraphics[width=70mm]{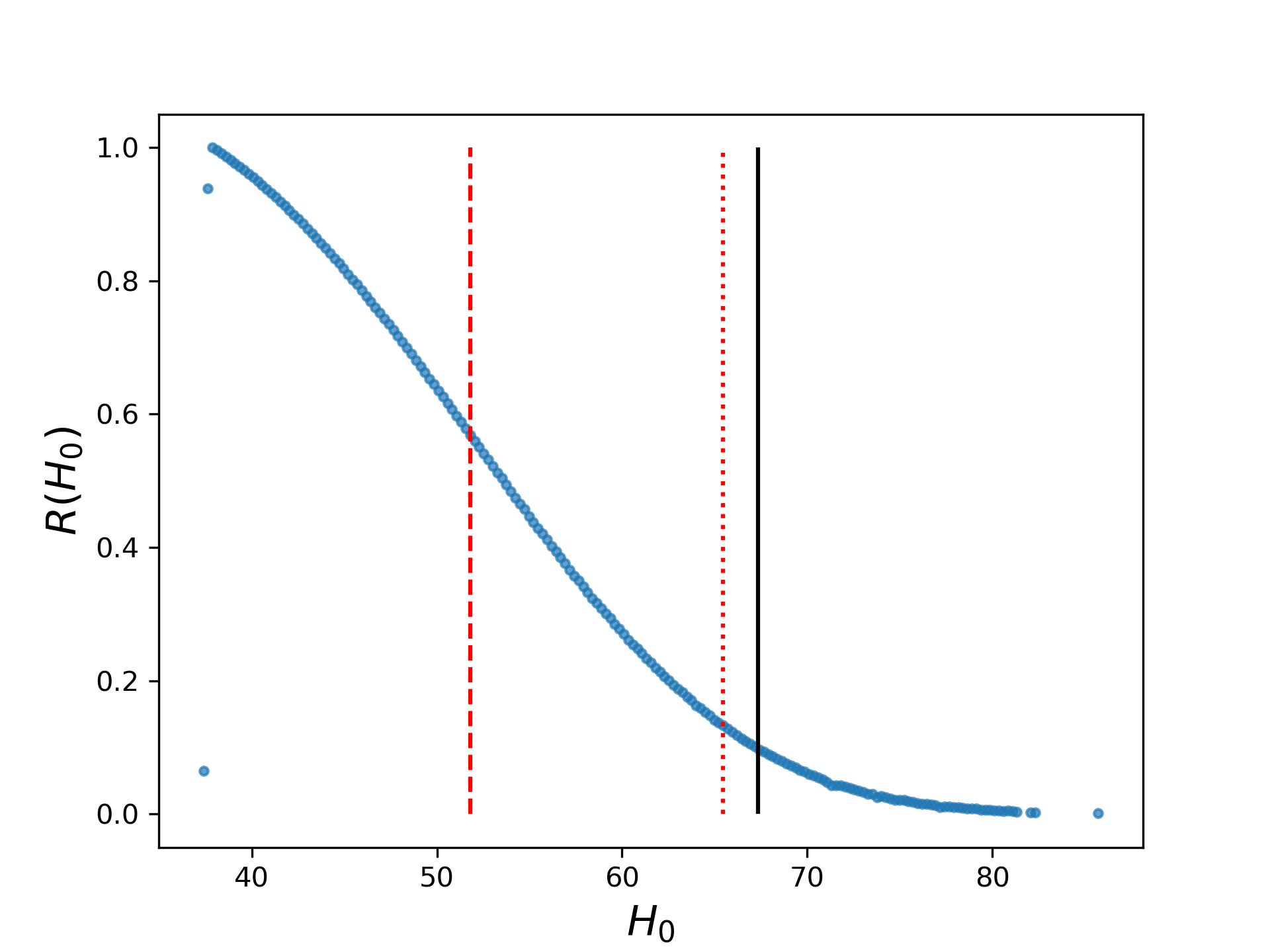} \\
\includegraphics[width=70mm]{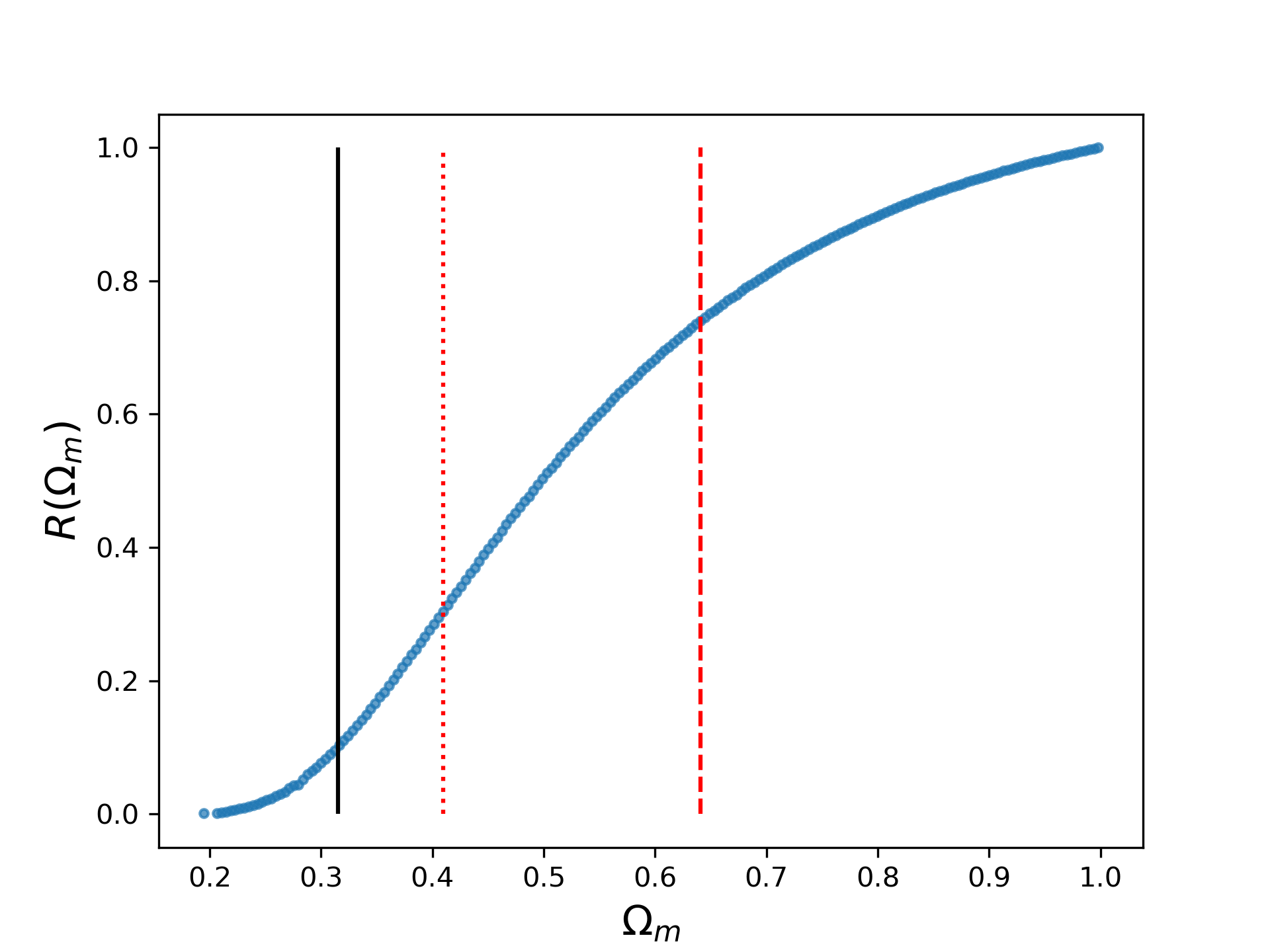}
\end{tabular}
\caption{$R(H_0)$ and $R(\Omega_m)$ distributions for OHD data with $z > 1.45$. Dashed and dotted red lines denote $68 \%$ ($1 \sigma$) and $95 \%$ ($2 \sigma$) confidence intervals. Black lines denote Planck best fit values. The two dots to the left of the $R(H_0)$ peak confirm that the distrbution goes to zero sharply below $H_0 = 40$ km/s/Mpc. In contrast, the $R(\Omega_m)$ distribution is one-sided and the peak is beyond $\Omega_m =1$.}
\label{fig:R_OHD_high} 
\end{figure}

{Before highlighting the tables confirming evolution outside of the errors, we begin with warmup exercises. In particular,} for OHD we focus on the $7^{\textrm{th}}$ row of Table \ref{tab:OHD}. While the table only focuses on a high redshift bin of varying redshift range, here we split the OHD sample of 54 data points into a low ($z < 1.45$) and high redshift ($ z \geq 1.45$) subsample of 47  and 7 data points, respectively. In Fig. \ref{fig:R_OHD_low} we show $R(H_0)$ and $R(\Omega_m)$ for the low redshift subsample. Evidently, the distributions are Gaussian, the Planck values (black lines) are within $2 \sigma$ (more accurately $1.5 \sigma$) and the distributions are peaked on values close to the best fits for the full sample. The latter is expected, as we have removed 7 high redshift data points from the full sample and the statistical weighting of these points is low. What our analysis in section \ref{sec:mock_data} shows is that low redshift data breaks a  degeneracy in the $(H_0, \Omega_m)$ parameters better than high redshift data, which means one expects larger errors from high redshift data, thus the lower statistical weighting.  In contrast, in Fig. \ref{fig:R_OHD_high} we confirm without using optimisation that the best fit parameters have shifted in the high redshift bin. In particular, the $R(\Omega_m)$ distribution is one-sided, implying the peak is beyond $\Omega_m = 1$, while the two points to the left of the peak in $R(H_0)$ tell us that lower values of $H_0$ are disfavoured. In general we take the lower bound on $H_0$ to be $H_0 \geq 0$, so there is no reason for the $R(H_0)$ distribution to terminate unless there is a sharp fall-off. The Planck values for $H_0$ and $\Omega_m$ are now excluded at $96 \%$ ($2.1 \sigma$) and $99\%$ ($2.6 \sigma$) confidence level, respectively.

\begin{figure}
   \centering
\begin{tabular}{c}
\includegraphics[width=70mm]{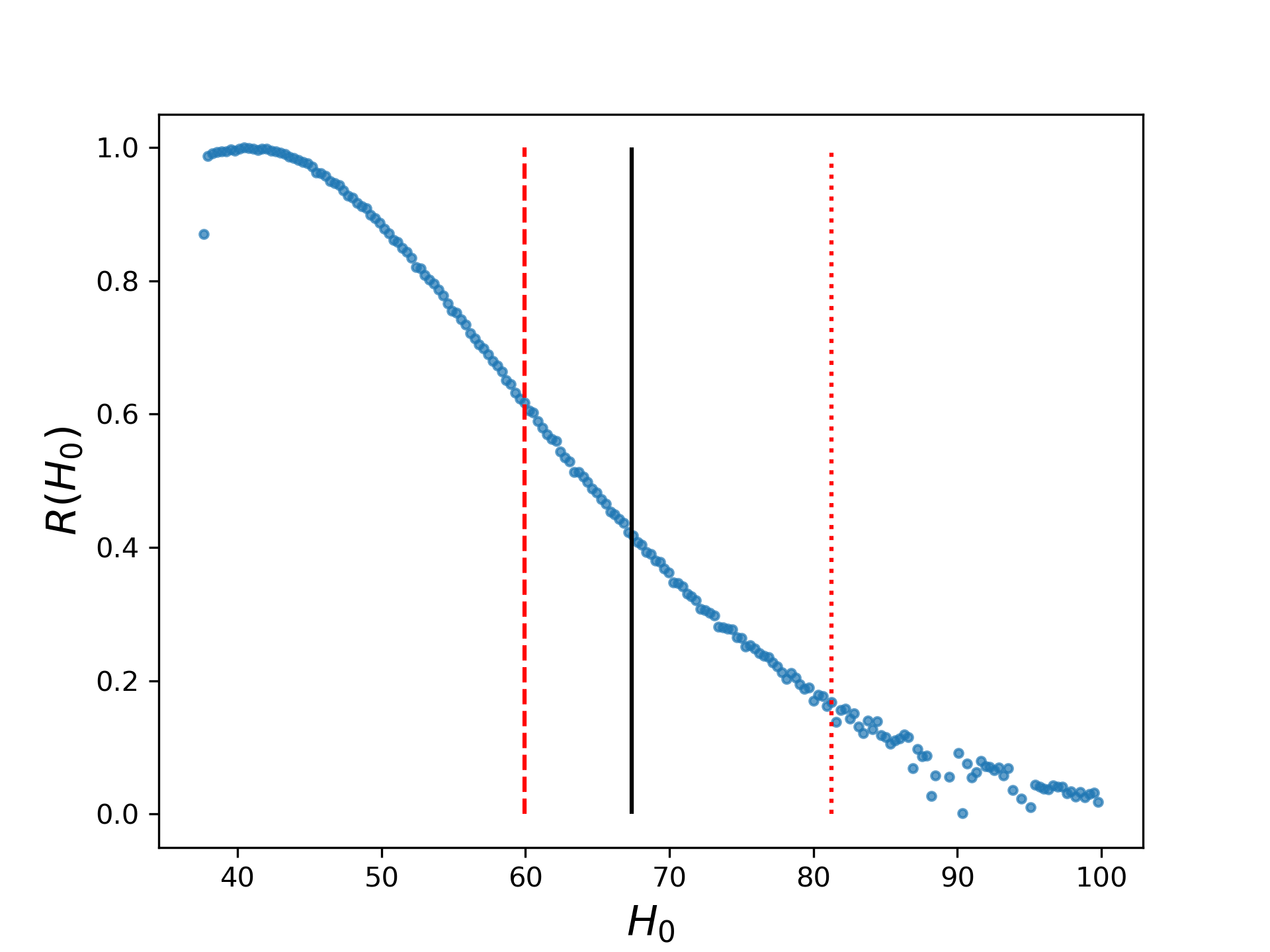} \\
\includegraphics[width=70mm]{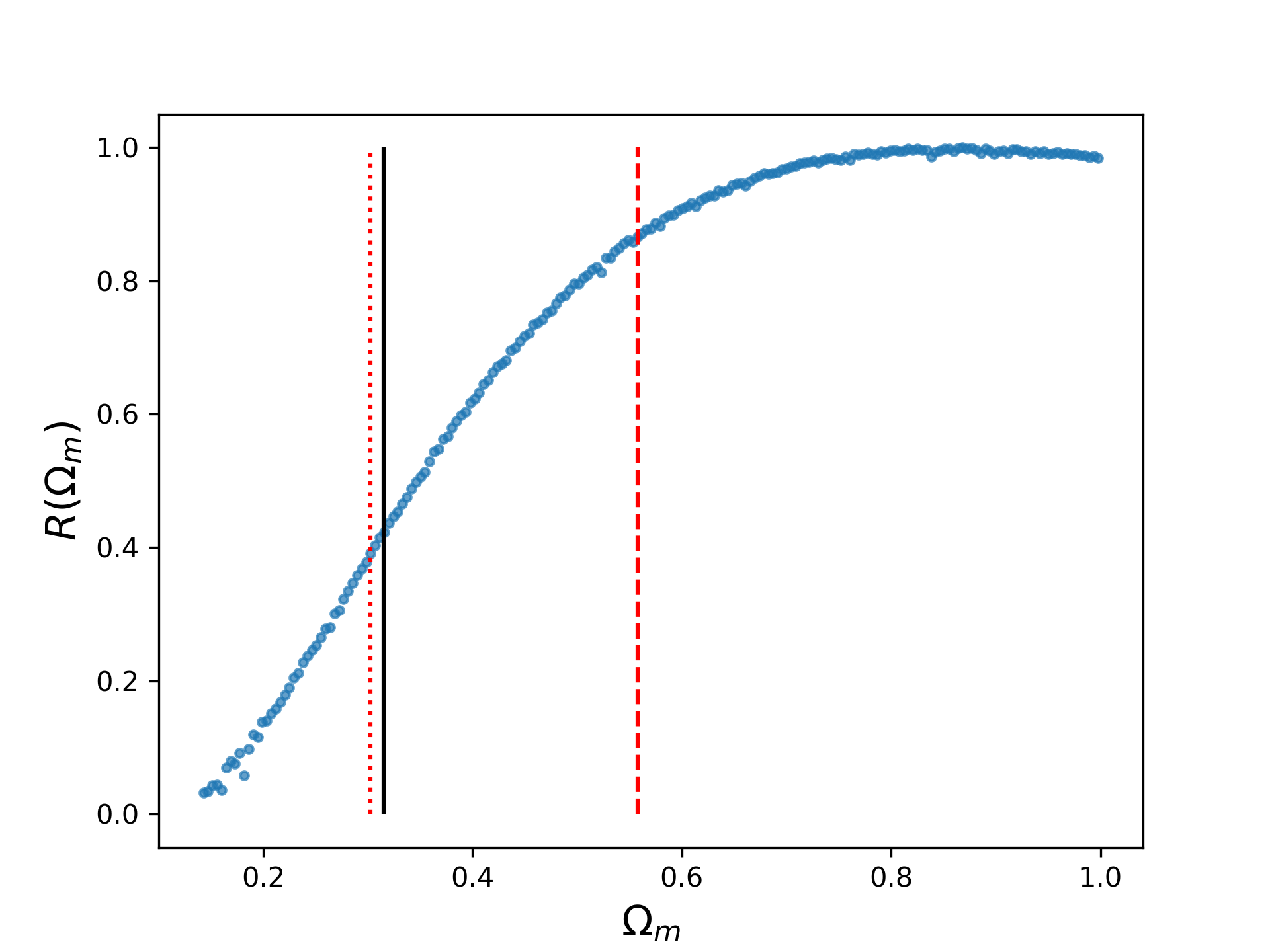}
\end{tabular}
\caption{$R(H_0)$ and $R(\Omega_m)$ distributions for Type Ia SNe data with $z > 0.95$. Dashed and dotted red lines denote $68 \%$ ($1 \sigma$) and $95 \%$ ($2 \sigma$) confidence intervals. Black lines denote Planck best fit values. The dot to the left of the $R(H_0)$ peak confirms it goes to zero below $H_0 = 40$ km/s/Mpc. A peak is evident in both $R(H_0)$ and $R(\Omega_m)$ distributions. Scatter is evident in the $R(H_0)$ distribution at higher $H_0$ values, but this can be removed by running a longer MCMC chain.}
\label{fig:R_SN_high} 
\end{figure}

This can be compared with $p = 0.021$ ($2.3 \sigma)$ from our OHD mock simulation in the same redshift range, and one recognises that the statistical significance is approximately the same. There is a slight difference in that our mock simulations are based on best fits for the full sample, and not the Planck values, but this is not expected to make a huge difference, especially since mock simulations assume a Planck prior on $\Omega_m h^2$. It is also compelling that the statistical significance from mock simulations, which treats the $(H_0, \Omega_m)$ parameters on par, is in the middle of the statistical significance inferred from $H_0$ and $\Omega_m$ profile distributions separately. It should be stressed that in the profile distribution analysis there is only one realisation of the data. Recovering the same statistical significance confirms that our mock simulations have not been impacted by anti-correlations (or degeneracies) between $H_0$ and $\Omega_m$ that may cause best fits from (noisy) mock simulations to move along the curves of constant $\Omega_m h^2$ in the $(H_0, \Omega_m)$-plane.

We next turn our attention to Type Ia SNe, where we focus on the redshift range corresponding to the $5^{\textrm{th}}$ entry in Table \ref{tab:SN}. As with OHD, we expect the low redshift $R(H_0)$ and $R(\Omega_m)$ distributions to be Gaussian, and consistent with Planck, so we focus exclusively on the high redshift segment $(z > 0.95)$. The MCMC chain now has an additional nuisance parameter $M_B$, which is relatively well constrained, so we do not discuss it further. In Fig. \ref{fig:R_SN_high} we show $R(H_0)$ and $R(\Omega_m)$ for high redshift SNe ($z > 0.95$). One can compare the peaks of the distribution to the best fits in Table \ref{tab:SN} ($5^{\textrm{th}}$ row) and confirm that $H_0$ and $\Omega_m$ distributions are peaked at $H_0 \sim 41$ km/s/Mpc and $\Omega_m \sim 0.86$ respectively. In contrast to the OHD data, the $R(\Omega_m)$ peak is noticeably not cut off by the $\Omega_m = 1$ bound. In the $R(H_0)$, there is some scatter in the $R(H_0)$ distribution beyond $H_0 = 80$ km/s/Mpc, but this can be removed by running a longer MCMC chain. The Planck values for $H_0$ and $\Omega_m$ are at $81 \%$ ($1.3 \sigma$) and $94 \%$ ($1.9 \sigma$) confidence level, respectively. This compares favourably with $p=0.081$ ($1.7 \sigma$) from our mock simulation analysis. Once again, our profile distribution analysis appears to average the statistical significances we see from $R(H_0)$ and $R(\Omega_m)$. 

\subsection{{Frequentist confidence intervals}}
{Having warmed up sufficiently, we will now use profile distributions to determine the missing errors in Tables \ref{tab:OHD}, \ref{tab:SN} and \ref{tab:QSO}. The results of this exercise are presented in Tables \ref{tab:OHD_errors}, \ref{tab:SN_errors} and \ref{tab:QSO_errors}. What the tables confirm is the following. Subject to the Planck prior on $\Omega_m h^2$, all data sets exhibit a decreasing $H_0$/increasing $\Omega_m$ trend with effective redshift, whereby best fits evolve outside of the errors. In particular, OHD data with $z \geq 1.2$ in Table \ref{tab:OHD_errors}, SNe data with $z \geq 0.9$ in Table \ref{tab:SN_errors} and QSO data with $z \leq 1$ in Table \ref{tab:QSO_errors} are already discrepant with Planck beyond the $68\%$ confidence level. Moreover, for OHD and SNe, high redshift subsamples disfavour the best fits of the full sample by in excess of $68\%$ confidence interval, whereas low redshift QSO subsamples disfavour the best fits of the full sample by in excess of $68\%$ confidence level. This establishes evolution outside of the errors in these three independent samples.}

\begin{table}
    \centering
    \begin{tabular}{c|cc}
         $z$ & $H_0$ (km/s/Mpc) & $\Omega_m$ \\
         \hline 
         \rule{0pt}{3ex} $0 \leq z \leq 2.36$ $(54)$ & $69.11_{-1.29}^{+1.29}$ &  $0.299_{-0.012}^{+0.012}$ \\
         \rule{0pt}{3ex} $0.5 \leq z \leq 2.36$ $(28)$ & $69.68_{-2.11}^{+2.03}$ &  $0.294_{-0.017}^{+0.019}$ \\
          \rule{0pt}{3ex} $0.7 \leq z \leq 2.36$  $(18)$ & $65.67_{-7.49}^{+6.66}$ &  $0.331_{-0.066}^{+0.105}$ \\
           \rule{0pt}{3ex} $1 \leq z \leq 2.36$ $(11)$ & $61.27_{-12.31}^{+10.56}$ &  $0.380_{-0.116}^{+0.282}$ \\
           \rule{0pt}{3ex} $1.2 \leq z \leq 2.36$ $(10)$ & $53.91_{-12.03}^{+10.07}$ &  $0.491_{-0.140}^{+0.310}$ \\
            \rule{0pt}{3ex} $1.4 \leq z \leq 2.36$ $(8)$ & $41.55_{-3.75}^{+13.62}$ &  $0.828_{-0.241}^{+0.169}$ \\
            \rule{0pt}{3ex} $1.45 \leq z \leq 2.36$ $(7)$ & $37.80_{-0.05}^{+14.04}$ &  $>0.641$ \\
             \rule{0pt}{3ex} $1.5 \leq z \leq 2.36$ $(6)$ & $37.80_{-0.23}^{+13.18}$ &  $>0.663$ \\
    \end{tabular}
    \caption{{Same as Table \ref{tab:OHD} but with $68\%$ confidence intervals determined through profile distributions. Note, at higher redshits the $\Omega_m$ confidence interval terminates at the bound $\Omega_m = 1$.}}
    \label{tab:OHD_errors}
\end{table}

\begin{table}
    \centering
    \begin{tabular}{c|cc}
         $z$ & $H_0$ (km/s/Mpc) & $\quad\Omega_{m}$ \\
         \hline
         \rule{0pt}{3ex} $0 < z \leq 2.26$ $(1048)$ & $69.26_{-2.32}^{+2.34}$ & $0.298_{-0.019}^{+0.022}$ \\
         \rule{0pt}{3ex} $0.7 < z \leq 2.26$ $(124)$ & $64.37_{-12.64}^{+13.92}$ & $0.345_{-0.106}^{+0.170}$  \\
         \rule{0pt}{3ex} $0.8 < z \leq 2.26$ $(82)$ & $58.99_{-12.63}^{+15.97}$ & $0.411_{-0.162}^{+0.276}$  \\
         \rule{0pt}{3ex} $0.9 < z \leq 2.26$ $(49)$ & $45.88_{-7.92}^{+13.91}$  & $0.679_{-0.190}^{+0.271}$  \\
         \rule{0pt}{3ex} $0.95 < z \leq 2.26$ $(34)$ & $40.73_{-3.14}^{+18.04}$ & $> 0.557$  \\
         \rule{0pt}{3ex} $1 < z \leq 2.26$ $(23)$ & $43.16_{-5.57}^{+22.66}$ & $> 0.503$ 
    \end{tabular}
    \caption{
    {Same as Table \ref{tab:SN} but with $68\%$ confidence intervals determined through profile distributions.}
    }
    \label{tab:SN_errors}
\end{table}

\begin{table}
    \centering
    \begin{tabular}{c|ccc}
         $z$ & $H_0$ (km/s/Mpc) & $\quad\Omega_{m}$ \\
         \hline 
         \rule{0pt}{3ex} $0 < z \leq 0.3$ $(56)$ & $406.41_{-138.48}^{+299.57}$ & $0.009_{-0.007}^{+0.080}$ \\
         \rule{0pt}{3ex} $0 < z \leq 0.5$ $(177)$ & $353.47_{-156.83}^{+61.03}$ & $ 0.011_{-0.003}^{+0.070}$ \\
         \rule{0pt}{3ex} $0 < z \leq 0.55$ $(233)$ & $433.91_{-200.47}^{+281.31}$ & $0.008_{-0.005}^{+0.021}$ \\
         \rule{0pt}{3ex} $0 < z \leq 0.6$ $(279)$ & $381.50_{-63.95}^{+283.99}$ & $0.010_{-0.007}^{+0.011}$ \\
         \rule{0pt}{3ex} $0 < z \leq 0.7$ $(398)$ & $73.40_{-22.41}^{+27.89}$ & $0.265_{-0.180}^{+0.278}$ \\
         \rule{0pt}{3ex} $0 < z \leq 0.8$ $(543)$ & $58.48_{-13.31}^{+21.81}$ & $0.418_{-0.194}^{+0.274}$ \\
         \rule{0pt}{3ex} $0 < z \leq 1$ $(826)$ & $40.69_{-2.75}^{+10.23}$ & $>0.625$ \\
          \rule{0pt}{3ex} $0 < z \leq 1.4$ $(1326)$ & $37.82_{-0.13}^{+9.35}$ & $>0.725$ \\
    \end{tabular}
    \caption{
    {Same as Table \ref{tab:QSO} but with $68\%$ confidence intervals determined through profile distributions.}}
    \label{tab:QSO_errors}
\end{table}

\subsection{Combining OHD and Type Ia SNe}
\begin{figure}
   \centering
\begin{tabular}{c}
\includegraphics[width=70mm]{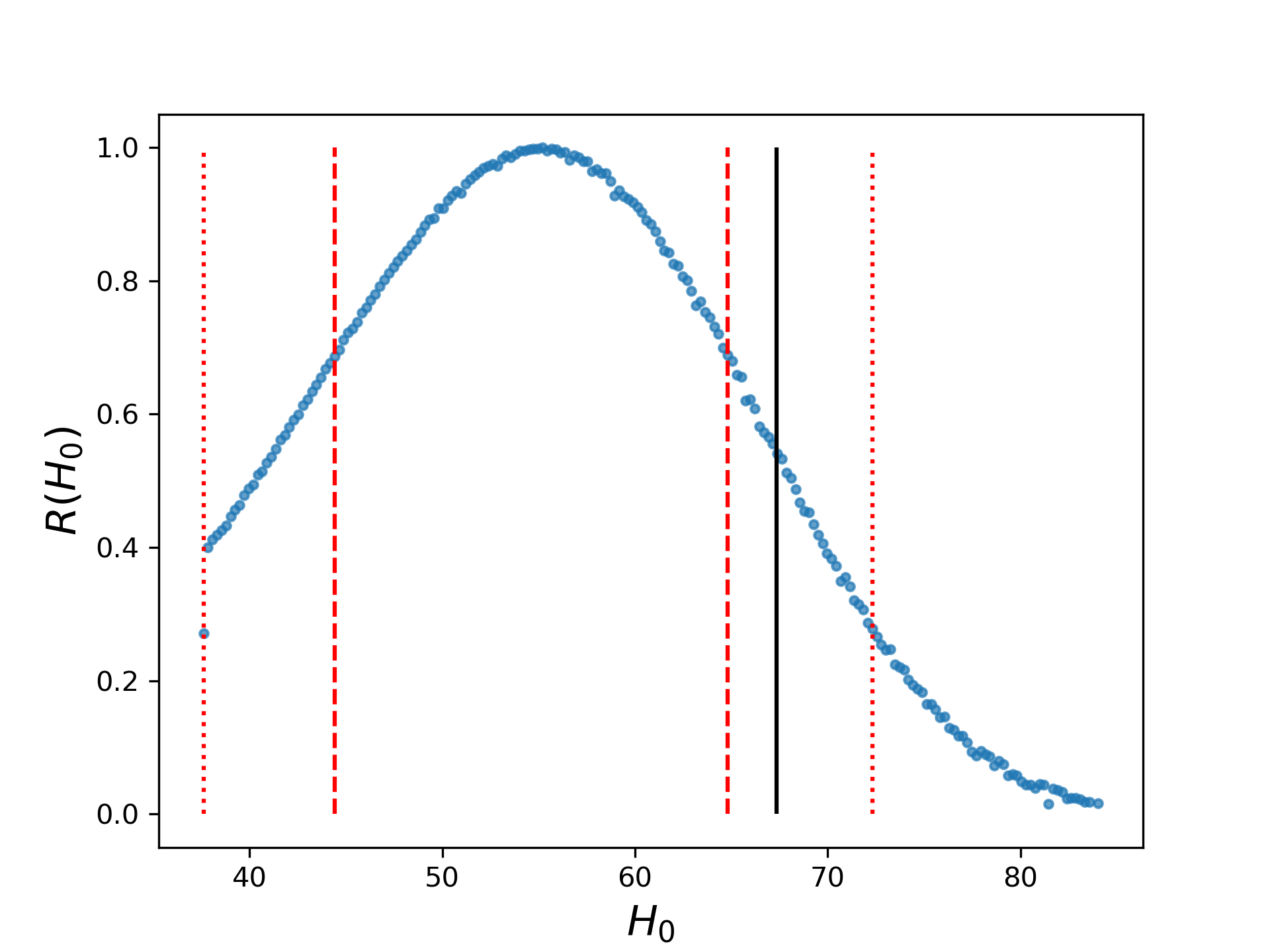} \\
\includegraphics[width=70mm]{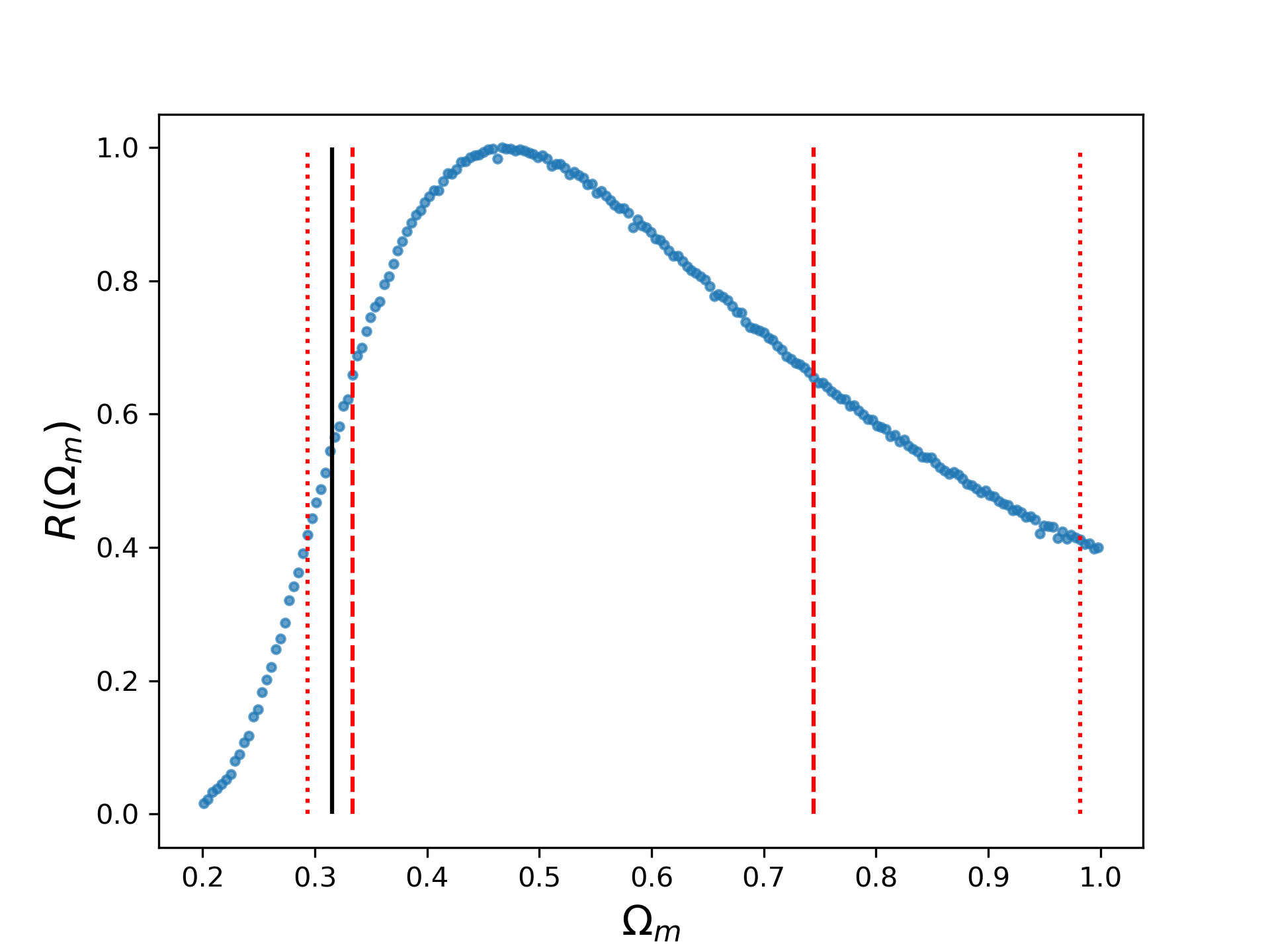}
\end{tabular}
\caption{$R(H_0)$ and $R(\Omega_m)$ distributions for a combination of OHD and Type Ia SNe data with $z > 0.95$. Dashed and dotted red lines denote $68 \%$ ($1 \sigma$) and $95 \%$ ($2 \sigma$) confidence intervals. Black lines denote Planck best fit values.}
\label{fig:R_OHD_SN_z95} 
\end{figure}

Lastly, since we see smaller values of $H_0$ and larger values of $\Omega_m$ in high redshift bins in {three} independent data sets, it is interesting to combine the likelihoods and repeat the exercise. {Here, we opt not to fold QSOs into the analysis, as they remain the most questionable observable. Nevertheless, if QSOs are standardisable, our results here are more significant than quoted.} We focus on two high redshift intervals $z > 0.95$ and $z > 1.45$, because these correspond to the redshift ranges where we see the most significant shifts in $(H_0, \Omega_m)$ in our mock simulations for Type Ia SNe and OHD, respectively. By comparing Tables \ref{tab:OHD} and \ref{tab:SN}, one sees that different data sets prefer different values of $H_0$ and $\Omega_m$ for data with $z \gtrsim 0.95$. As a result, one expects an average value, and this is indeed what we find in our profile distribution analysis. In Fig. \ref{fig:R_OHD_SN_z95} we show the result. We note a shift in the peak to smaller $H_0$ and larger $\Omega_m$ values, but the significance is not so great, in the sense that the Planck values are now excluded at $81 \%$ ($1.3 \sigma$) confidence level. This is more significant than $p = 0.258$ ($0.7 \sigma$) one sees in mock simulations of OHD in a comparable redshift range ($4^{\textrm{th}}$ row of Table \ref{tab:OHD}). Evidently, the addition of SNe data to OHD pulls $H_0$ lower, pulls $\Omega_m$ higher and increases the statistical significance.

\begin{figure}
   \centering
\begin{tabular}{c}
\includegraphics[width=70mm]{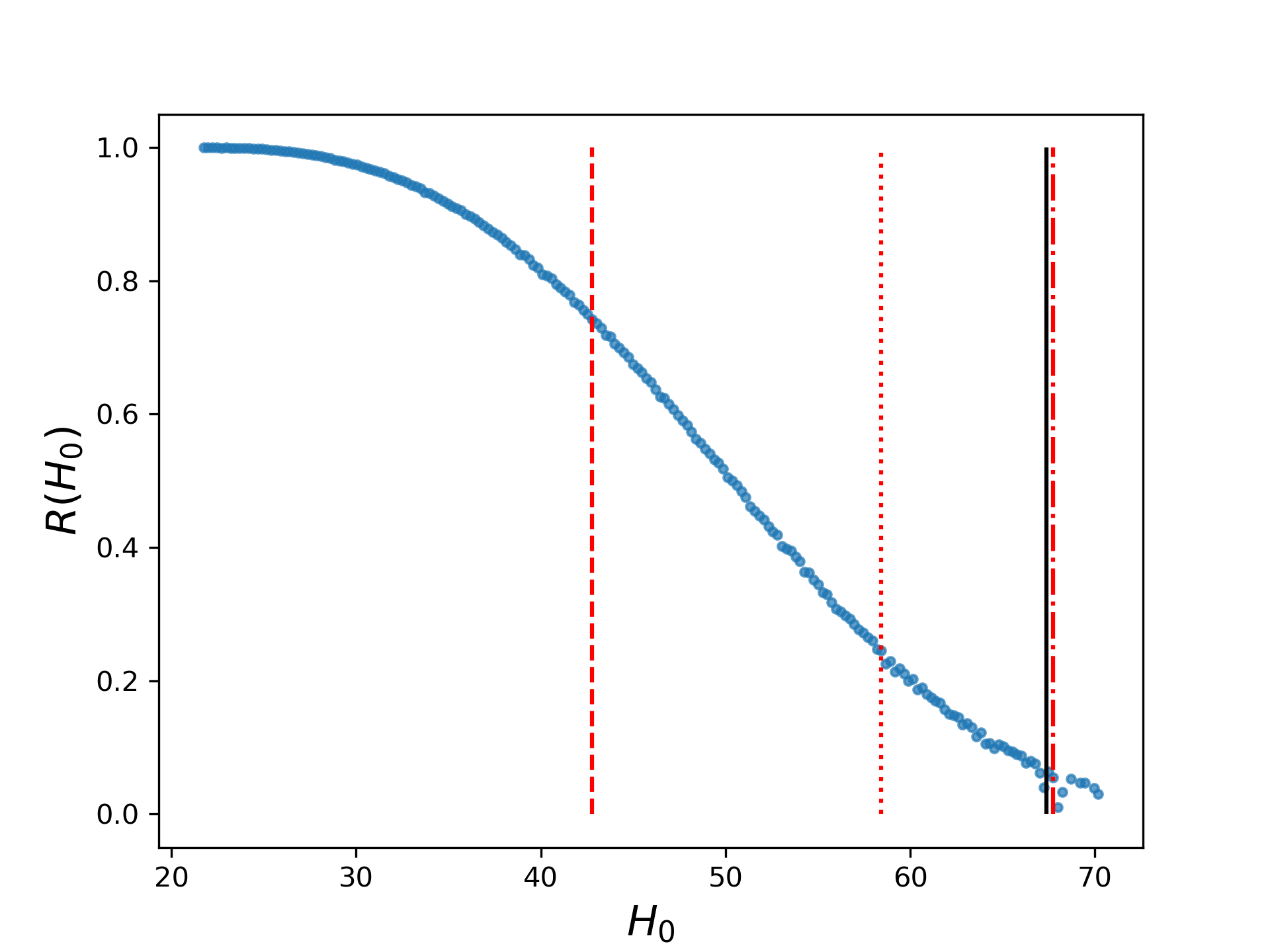} \\
\includegraphics[width=70mm]{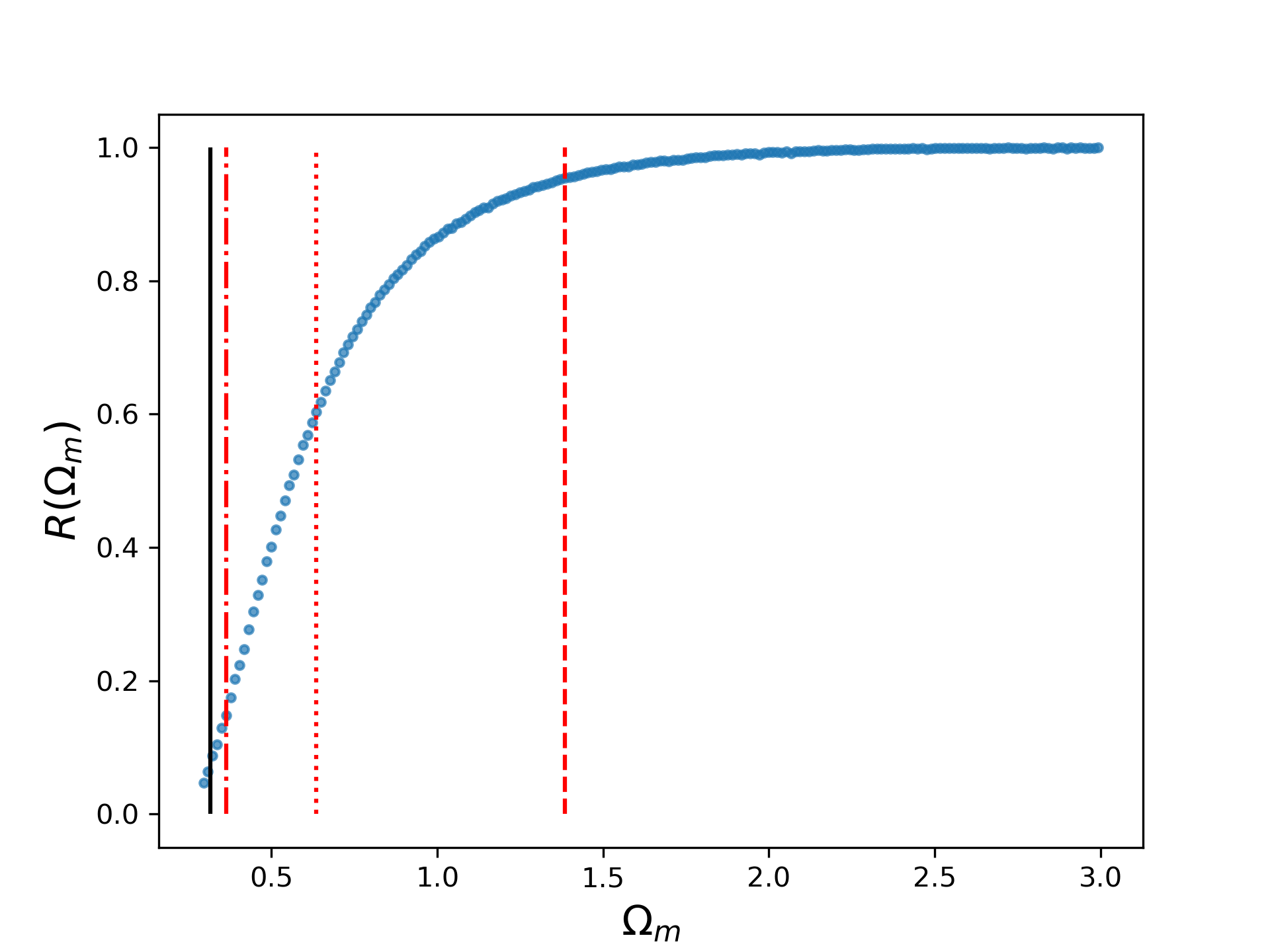}
\end{tabular}
\caption{$R(H_0)$ and $R(\Omega_m)$ distributions for a combination of OHD and Type Ia SNe data with $z > 1.45$. Dashed, dotted and dashed-dotted red lines denote $68 \%$ ($1 \sigma$), $95 \%$ ($2 \sigma$) and $99.7 \%$ ($3 \sigma$) confidence intervals. Black lines denote Planck best fit values. We have relaxed the traditional $\Omega_m \leq 1$ prior to $\Omega_m \leq 3$ in order to include the peak at $\Omega_m \sim 2.8$. Although no points are evident to the left of the $R(H_0)$ peak, the distribution falls off to zero as we imposed an $H_0 \geq 0$ prior.}
\label{fig:R_OHD_SN_z145} 
\end{figure}

Our final exercise is to combine OHD and SNe with $z > 1.45$. It should be stressed that SNe are extremely sparse at these redshifts with only 6 Pantheon SNe in the range. As a result, we can expect OHD to have greater bearing on the outcome. As is clear from Table \ref{tab:OHD} and Fig. \ref{fig:R_OHD_high}, the $\Omega_m \leq 1$ prior precludes the data from finding the point in $\Lambda$CDM parameter space that best fits the data (minimum of $\chi^2$). Thus, here we relax the $\Omega_m$ prior enough to $\Omega_m \leq 3$ to find the best fit $\Omega_m$ value at $\Omega_m \sim 2.8$. The resulting $R(H_0)$ and $R(\Omega_m)$ distributions are shown in Fig. \ref{fig:R_OHD_SN_z145}. The key take away from these plots is that the $R(H_0)$ and $R(\Omega_m)$ distributions preclude the Planck value at $99.6 \%$ ($2.9 \sigma$) and $99.97 \%$ ($3.6 \sigma$) confidence level, respectively. As our mock analysis in section \ref{sec:mock_data} has shown (see also \cite{Colgain:2022tql}), there is nothing to preclude $\Omega_m > 1$ best fits in high redshift bins assuming the $\Lambda$CDM model, since this can happen in mock data based on Planck values. The physical regime corresponding to $\Omega_m \leq 1$ is excluded at $ 83 \%$ ($1.4 \sigma$) confidence level.

\section{Conclusions}
We explained through analytic arguments and simulations why the Planck value $\Omega_m \sim 0.3$ is less likely when one fits higher redshift binned $H(z)$ observations to the flat $\Lambda$CDM model. Our arguments are independent of mock input parameters and simply follow from the irrelevance of the $A$ term in \eqref{LCDM} at higher redshifts, and $A\geq 0$, which yield an initial `pile up' of best fits on $\Omega_m = 1$, before piling up at $\Omega_m \sim 0$ at even higher redshifts.  This reduces the probability of recovering the Planck value at high redshift, thus providing an avenue to test the model. Note, it is not enough to find unexpected best fits, but one must prove that those best fits are statistically unlikely assuming no evolution of $\Lambda$CDM parameters across data sets. Alternatively put, one must first establish evolution in best fit cosmological parameters before addressing errors and statistical significance, since if there is no evolution in best fit parameters, the size of the errors makes little difference.
Our warm-up DESI mock analysis here solely pertains to $H(z)$ constraints, but the same conclusions hold for angular diamater distance $D_{A}(z) \propto \int_{0}^{z} \dd z^{\prime}/H(z^{\prime})$ constraints \cite{Colgain:2022tql}. The reader will note that $D_{L}(z) \propto D_{A}(z)$, so all our observed data is in the $H(z)$ or $D_{L}(z)/D_{A}(z)$ class. {We emphasise again that the role of the first section is simply to explain why the mocks in Figs. \ref{fig:OHDsim}, \ref{fig:SNsim} and \ref{fig:QSOsim} are non-Gaussian, but this has no bearing on the $p$-values.}

In the second part of our work, we confirmed a decreasing $H_0$/increasing $\Omega_m$ behaviour in OHD and Type Ia SNe with $p$-values as low as $p = 0.021$ ($2.3 \, \sigma$) and $p=0.081$ ($1.7 \, \sigma$), respectively. We resorted to comparison to mock analysis in the same redshift range with the same number of data points and same errors because i) mock simulations are traditionally how one approaches anomalies in data, e. g. \cite{Eriksen:2003db}, and ii) one can circumvent the difficulty estimating errors with non-Gaussian distributions. Using Fisher's method, the combined (lowest) probability for these established cosmological probes is $p = 0.013$ ($2.5 \, \sigma$). In QSOs, an intrinsically high redshift emerging observable, we see the opposite trend where discrepant best fit $(H_0, \Omega_m)$ values relative to the entire sample appear at lower redshifts with probabilities as low as $p = 0.019$ ($2.3 \, \sigma$). Once again combining the probabilities, one finds a (lowest) probability $p = 0.0021$ ($3.1 \, \sigma$). One may also benchmark with respect to the bin $0 < z \leq 0.7$, where $\Omega_m \sim 0.3$, in which case the combined (lowest) probability becomes $p = 0.0078$ ($2.7 \, \sigma$). As argued in the text, the probabilities we quote are lower bounds, especially since we preclude $\Omega_m > 1$ in observed and mock data fitting. One can of course find redshift ranges with less evolution, but if there is a \textit{bona fide} decreasing $H_0$/increasing $\Omega_m$ best fit trend with increasing effective redshift, then one expects decreasing $p$-values in Tables \ref{tab:OHD}-\ref{tab:SN}. Moreover, by working with the full sample and not binning it, one can return to the working \textit{assumption} that there is no evolution in the samples. Our analysis here challenges the working assumption. 

Since mock simulations may be unfamiliar to some readers, we revisited the redshift ranges where we see the greatest shift away from canonical Planck cosmological parameters with profile distributions. {Our analysis focused on the better understood observables, namely OHD and SNe, but we presented $68\%$ confidence intervals for all observables across different redshift ranges, allowing us to confirm that the best fit parameters evolve outside of the errors. Profile distributions} allow us to recycle the information in the MCMC chain directly, find the most probable values for cosmological parameters and establish the confidence intervals at which the Planck ($H_0, \Omega_m)$ values are excluded in high redshift bins. Throughout we find good agreement between mock simulations and profile distributions analysis. In particular, we find that the statistical significance from mock simulations averages the statistical significance we see in $H_0$ and $\Omega_m$ profile distributions. 
The main take away is that high redshift OHD and Type Ia SNe samples prefer larger values of $\Omega_m$ and smaller values of $H_0$ and this preference is statistically significant at $\sim 3 \sigma$ when we relax the $\Omega_m \leq 1$ prior. The conclusion here is separately confirmed in re-analysis with later and better quality data \cite{Malekjani:2023dky, Colgain:2023bge}. Throughout it should be stressed that we are looking at small subsamples of OHD and Type Ia SNe data sets at high redshifts, but the discrepancies we see are statistically significant. {Given the small size of the samples, it is imperative to revisit results as data quality improves further. Note, QSOs have no problems with statistics, but systematics may be an issue. There is a statistically significant discrepancy with Planck at higher redshifts $ z \gtrsim 1.5$ reported 
 elsewhere in the literature \cite{Risaliti:2018reu}.}

Objectively, all observables show signatures of evolution to lower $H_0$ values and higher $\Omega_m$ values between low and high redshifts, in line with mock expectations that they can be easily displaced from Planck values.  {Neglecting selection effects and more general systematics across multiple observables (SNe, cosmic chronometers, BAO, QSOs)}, this supports the idea that the flat $\Lambda$CDM model is a dynamical model where fitting parameters, which should be constants, evolve in (cosmic) time. This cautions that cosmological tensions may be an outcome of the flawed assumption that $(H_0, \Omega_m)$ are unique within flat $\Lambda$CDM. {In short, the $\Lambda$CDM model appears to have broken down; no model of physical interest should make different predictions at different epochs. If substantiated, this settles the systematics versus missing physics debate on $\Lambda$CDM tensions. The outcome may not be so surprising. What is being tested is, given the current quality of high redshift $z \gtrsim 1$ data, whether exclusively high redshift data can recover the Planck values. Our findings are apparently no, but this conclusion may be reversed as high redshift data improves, thereby throwing a lifeline to a $\Lambda$CDM model besmirched by persistent tensions.}

{Finally we note that one could attempt to interpret these findings in terms of an underdensity in the Universe at the scale of a few Gpc, e. g. \cite{Ding:2019mmw, Haslbauer:2023vyf}.\footnote{{See also \cite{Shanks:2018rka, Kenworthy:2019qwq, Haslbauer:2020xaa, Cai:2020tpy, Camarena:2022iae} for claims and counterclaims regarding the ability of smaller voids to resolve $H_0$ tension.}} However, before doing so, it is imperative to check if matter is \textit{observationally} pressureless \footnote{{Theoretically, it is pressureless, but observation and theory need not agree.}}. Note, the (flat) $\Lambda$CDM model is so simple that DE and matter sectors are coupled through a single parameter $\Omega_m$, and it is prudent to confirm that $\Omega_m h^2$ is not evolving in high redshift bins before making further deductions. More concretely, one needs to check that the Hubble parameter scales as $H(z) \sim 100 \sqrt{\Omega_m h^2} (1+z)^b$ with constant $\Omega_m h^2$ and $b = \frac{3}{2}$ in high redshift bins. If either constant evolves, this contradicts the assumption that matter is pressureless.}

\section*{Acknowledgements}
We thank Stephen Appleby, Eleonora Di Valentino, Adam Riess, Joe Silk and Jenny Wagner for correspondence. We also thank Bum-Hoon Lee, Wonwoo Lee, Albin Nilsson, Mehrdad Mirbabaei, Somyadip Thakur and Lu Yin for discussion. E\'OC was supported by the National Research Foundation of Korea grant funded by the Korea government (MSIT) (NRF-2020R1A2C1102899). DS is partially supported by the US National Science Foundation, under Grant No. PHY-2014021. MMShJ would like to acknowledge SarAmadan grant No. ISEF/M/400121 and ICTP HECAP section where this project was carried out. M.G. acknowledge the support of Division of Science at NAOJ. This article/publication is based upon work from COST Action CA21136 – “Addressing observational tensions in cosmology with systematics and fundamental physics (CosmoVerse)”, supported by COST (European Cooperation in Science and Technology).

\appendix

\section{Removing $\Omega_m h^2$ prior}
\begin{figure}
   \centering
\begin{tabular}{c}
\includegraphics[width=70mm]{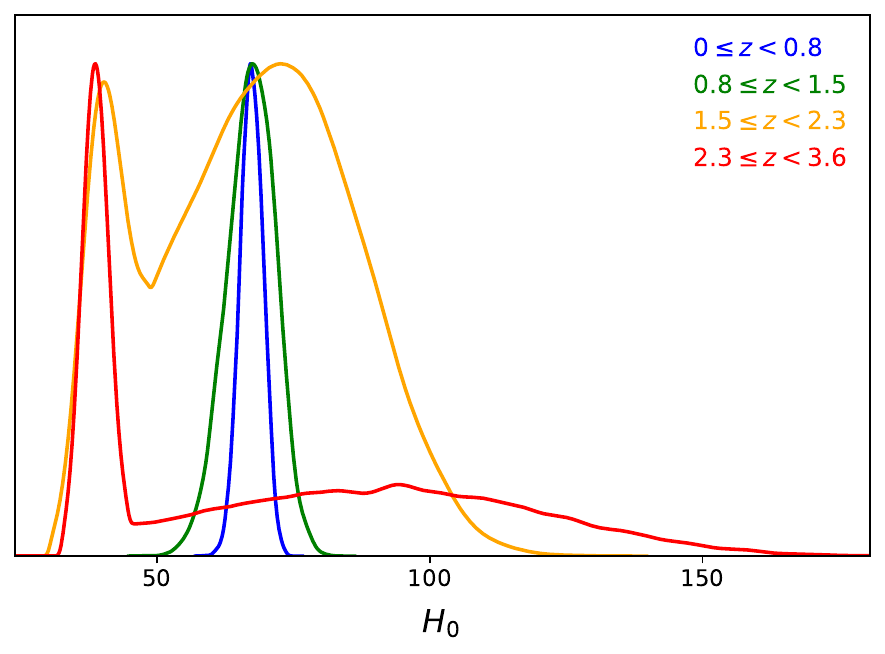} \\
\includegraphics[width=70mm]{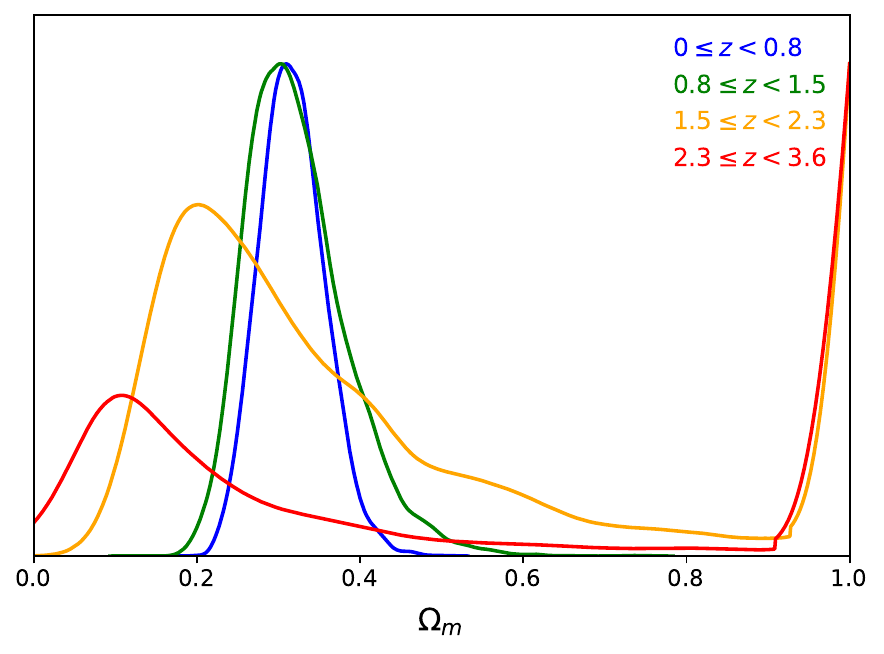}
\end{tabular}
\caption{Same as Fig. \ref{fig:H0Om_dist} but without the $\Omega_m h^2$ prior.}
\label{fig:H0Om_no_prior_dist} 
\end{figure}

Removing the Planck $\Omega_m h^2$ prior from Fig. \ref{fig:H0Om_dist} leads to a spreading in all distributions, but qualitatively the features are the same, as expected from the analytic discussions. This can be confirmed in Fig. \ref{fig:H0Om_no_prior_dist}.  

\begin{figure}
   \centering
\includegraphics[width=80mm]{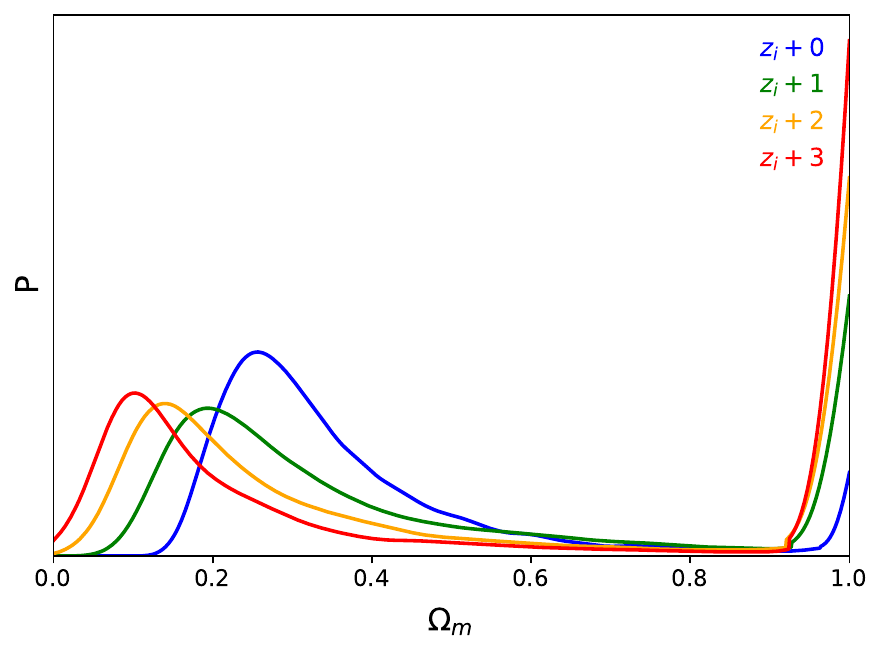}
\caption{Probabilities of a given $\Omega_m$ best fit value for forecast DESI $H(z)$ data in the range $2.3 \leq z < 3.6$ (blue curve) with Planck input values. Green, yellow and red denote the probabilities if the same data are displaced to higher redshifts.}
\label{fig:DESI_displaced} 
\end{figure}

\section{Confirmation of $P(\Omega_m \sim 0.3) \rightarrow 0$}
In this section we consider the same mocking procedure but focus exclusively on the fourth DESI bin with redshift range $2.3 \leq z < 3.6$. We now displace the redshifts in intervals of $+1$ without changing the data and document the effect on the distribution of $\Omega_m$ best fits over a few thousand mocks. Noting that very high redshift $H(z)$ data only constrains $\Omega_m h^2 \propto B$ well, we expect $H_0$ and $\Omega_m$ to be largely unconstrained, corresponding to flat/uniform distributions in $H_0$ and $\Omega_m$ at high redshift. In particular, we expect $\Omega_m$ distributions to flatten as the effective redshift increases. Here we confirm that this flattening happens through a shift in the peak of the (non-Gaussian) $\Omega_m$ distribution to smaller values away from $\Omega_m = 0.3$, so that the probability of encountering an $\Omega_m$ best fit close to canonical values $P(\Omega_m \sim 0.3)$ decreases.

Once again, we assume Planck input parameters, $H_0 = 67.36, \Omega_m = 0.315$ and the Gaussian prior, $\Omega_m h^2 = 0.1430 \pm 0.0011$ \cite{Planck:2018vyg}. In Fig. \ref{fig:DESI_displaced} we present (normalised) probabilities for $\Omega_m$ for Planck-$\Lambda$CDM mocks, where the blue curve corresponds to the red curve in Fig. \ref{fig:H0Om_dist}. The remaining curves correspond to $\Omega_m$ probabilities as we displace the original binned data in redshift. Since we are at high redshift, the probability of $\Omega_m=1$, $P(\Omega_m=1) > 0$ and it clearly increases with redshift of the sample. This is evident from the greater number of best fits piled up at $\Omega_m = 1$. The shift in the peak of $\Omega_m$ best fits is a result of the best fits stretching along a constant $\Omega_m h^2$ curve in the ($H_0, \Omega_m)$-plane. This leads to a larger number of configurations at smaller $\Omega_m$ values and a shift in the peak in the $\Omega_m$ distribution to smaller values when projected onto the $\Omega_m$ axis.
 Fig. \ref{fig:DESI_displaced} shows that this trend is more pronounced at higher redshifts and it is an obvious implication that at a given high redshift, any knowledge of the input parameters is lost and the probability of recovering the Planck value, $P(\Omega_m \sim 0.3)$ is close to zero. Potentially other $\Omega_m$ values favoured, as can be seen from our mocks.

\begin{figure}
   \centering
\includegraphics[width=80mm]{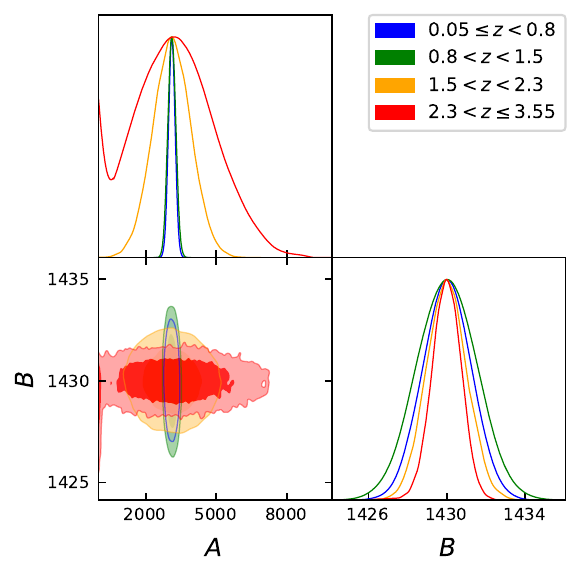}
\caption{A corner plot with DESI forecast data demonstrating that $(A, B)$ are uncorrelated across the bins.}
\label{fig:AB_uncorrelated} 
\end{figure}

\section{Further comments on $(A,B)$}
In this section we show in Fig. \ref{fig:AB_uncorrelated} that the derived (secondary) parameters $(A,B)$ are uncorrelated. As explained in the main text, we imposed a (strong) Planck Gaussian prior on $B$, so unsurprisingly $B$ conforms to a Gaussian and $A$ is also Gaussian where it is not impacted by the bound $\Omega_m \leq 1$. Noting that $(A,B)$ are uncorrelated, whereas the transformation from the fitting parameters $(H_0, \Omega_m)$ to $(A, B)$ is non-linear, it would be surprising if one encountered Gaussian distributions in all parameters.

\begin{figure}
   \centering
\includegraphics[width=80mm]{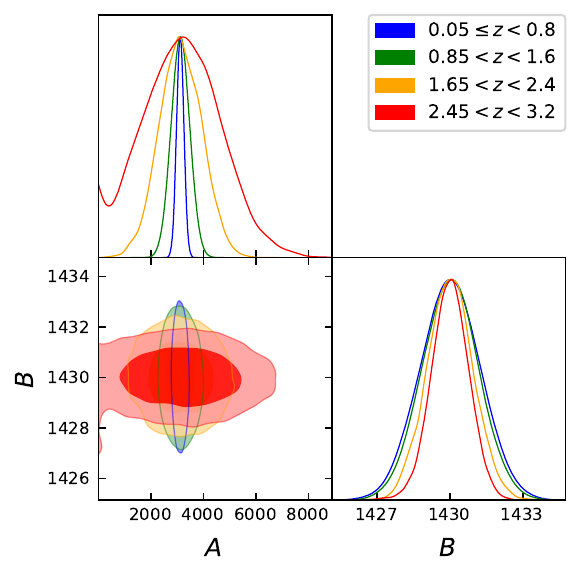}
\caption{A corner plot with the same data in all bins. The $A$ distribution spreads whereas the $B$ distribution narrows once one has the same data in all bins.}
\label{fig:AB_uniform_quality} 
\end{figure}

There is another loose end to close. The astute reader will notice that the $B$ distribution does not narrow uniformly in Fig. \ref{fig:AB_uncorrelated}. To explain this feature we note that the percentage $H(z)$ errors vary with redshift in the DESI forecast \cite{DESI:2016fyo} and the errors are smallest at the boundary of bin 1 and bin 2 ($z \sim 0.8$). As a result, bin 1 better constrains $B$, the relevant high redshift parameter, whereas bin 2 better constrains $A$, the relevant low redshift parameter. From Fig. \ref{fig:AB_uncorrelated} one notes that any spread in $A$ is marginal between bins 1 and 2, while the $B$ distribution actually spreads between bins 1 and 2, thus contradicting statements in the text. However, in Fig. \ref{fig:AB_uniform_quality} we produce four bins with \textit{exactly} the same data in each bin by simply displacing the percentage errors in bin 1 in redshift and using them as the basis for mocks in bins 2, 3 and 4. As a result, one has the same percentage errors in each bin, and one notices that $A$ spreads whereas $B$ narrows with effective redshift. In summary, one generically expects a spreading $A$ distribution and narrowing $B$ distribution with effective redshift in the flat $\Lambda$CDM model as the model transitions from a two-parameter model to an effective one-parameter model, but this trend may be impacted by the number of data points and the errors.

\section{MCMC analysis}
As explained in the text, great care is required with MCMC inferences in non-Gaussian regimes. For this reason we opted to compare our best fits in observed data directly with best fits in mock data \textit{in the same redshift range with the same number of data points and errors}. Here we take a look at the inferences one would make with MCMC, where we focus on OHD data, since it is the simplest to analyse because there are no nuisance parameters. Once again, we impose the bounds $0 \leq \Omega_m \leq 1$. We then split the OHD sample at $z = 1.45$, which is of interest since it corresponds to the $7^{\textrm{th}}$ entry in Table \ref{tab:OHD}, where we have recorded the lowest probability of recovering best fits from observed data in mock data. It should be noted that our best fit $\Omega_m$ value saturates the bound $\Omega_m =1$, so we expect that our MCMC distribution has no $\Omega_m$ peak because it is precluded by the priors. 

\begin{figure}
   \centering
\includegraphics[width=80mm]{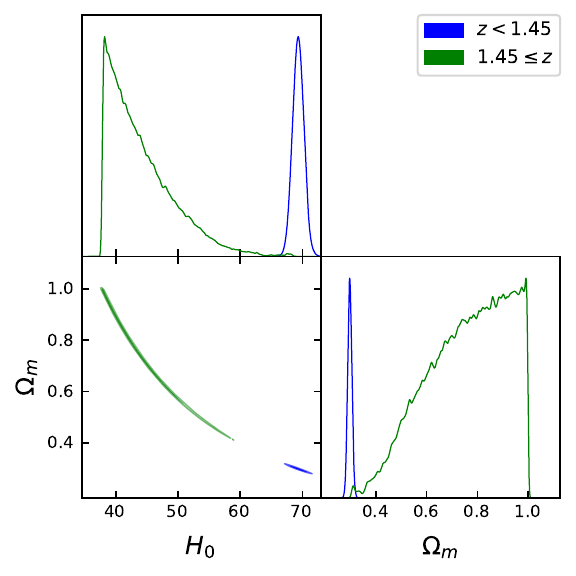}
\caption{Inferences of cosmological parameters from MCMC chains in the OHD sample of 54 data points. Evolution of cosmological parameters is evident when comparing low and high redshift subsamples.}
\label{fig:OHD_lowz_highz} 
\end{figure}

In Fig. \ref{fig:OHD_lowz_highz} we show the outcome of the MCMC analysis with \textit{GetDist} \cite{Lewis:2019xzd}. As expected, the low redshift ($z < 1.45$) sample of 47 OHD data points leads to an $\Omega_m$ distribution that is perfectly Gaussian, but the high redshift ($ 1.45 \geq z$) sample of 7 OHD data points does not. The $\Omega_m$ distribution continues to increase towards the $\Omega_m=1$ bound implying either a peak at the bound or beyond the bound. Priors are clearly impacting the result. In the $(H_0, \Omega_m)$-plane the contours follow a curve of constant $\Omega_m h^2$ due to the Planck prior. This curve is elongated in the high redshift sample and the discrepancy between the low and high redshift subsamples of the full OHD sample is evident in the $(H_0, \Omega_m)$-plane. From the MCMC chains, we infer the constraints on $(H_0, \Omega_m)$ from the low redshift sample to be $(H_0, \Omega_m) = (69.32^{+0.90}_{-0.90}, 0.298^{+0.008}_{-0.008})$, whereas the constraints from the high redshift sample are $(H_0, \Omega_m) = (42.82^{+6.56}_{-3.69} ,0.779^{+0.154}_{-0.194})$. Here, we quote $16^{\textrm{th}}, 50^{\textrm{th}}$ and $84^{\textrm{th}}$ percentiles in line with standard practice, of course assuming a Gaussian distribution. It is clearly wrong to do this as our distributions are non-Gaussian and have been impacted by the $\Omega_m \leq 1$ bound, but the results are merely indicative of smaller $H_0$ values/larger $\Omega_m$ values at higher redshifts. To be clear, it should be evident that relaxing the $\Omega_m \leq 1$ prior will allow the 2D posterior to stretch further into the top left corner of the $(H_0, \Omega_m)$-plane. This will shift the peak of the $H_0$ posterior to smaller values, thereby exacerbating tensions in projected 1D $H_0$ posteriors (see Fig. 2 of \cite{Malekjani:2023dky}). Once the MCMC posteriors stretch to the extent that they become prior dependent, there are very few robust conclusions one can draw.

\end{document}